 \documentstyle[11pt,aaspp4]{article}  

\slugcomment{ApJ, Version, July 20, 2000}

\lefthead{Fenimore \& Ramirez-Ruiz}
\righthead{Redshifts for BATSE Gamma-Ray Bursts}
\begin{document}
\title{Redshifts For 220 BATSE Gamma-Ray Bursts Determined by
Variability and the Cosmological Consequences} 
\author{E.~E.~Fenimore$^{1}$, and E.~Ramirez-Ruiz$^{1,2}$}  
\affil{$^1$MS D436, Los Alamos National Laboratory, Los Alamos, NM 87545}
\affil{$^2$Institute of Astronomy, University of Cambridge, 
Cambridge, CB3 0HA United Kingdom}

\def\domega{{\rm d}$\Omega$}
\def\lognp{Log~$N$--Log~$P_{256}$}
\def\lognl{Log~$N$--Log~$L_{256}$}
\def\Mesz{M\'esz\'aros}
\def\lap{\hbox{{\lower -2.5pt\hbox{$<$}}\hskip -8pt\raise-2.5pt\hbox{$\sim$}}}
\def\gap{\hbox{{\lower -2.5pt\hbox{$>$}}\hskip -8pt\raise-2.5pt\hbox{$\sim$}}}
\def\ul#1{$\underline{\smash{\vphantom{y}\hbox{#1}}}$}
\begin{abstract}

We show that the time variability of a 
gamma-ray burst (GRB)  appears to be 
correlated with the absolute luminosity of the burst: smooth bursts are
intrinsically less luminous. This
Cepheid-like relationship
can be used to determine the
redshift of a GRB from parameters measured solely at gamma-ray
energies.  
The
relationship is based on only seven events at present and needs to be
further confirmed with
more events. 
We present the details of converting GRB observables to 
luminosities and redshifts for 220 bright, long GRBs from the Burst
and Transient Source Experiment (BATSE)
and explore the cosmological consequences.
In particular, we derive  the GRB  rate as a function of $z$ without
assuming either a luminosity function
or that the GRB rate follows the star formation rate (SFR).
We find that the GRB formation rate scales
as $(1+z)^{3.3 \pm 0.3}$.
The observations used to derive the SFR can be strongly affected by dust for
$z \gap 2$
whereas  GRB
observations are not. If GRBs trace  star formation, then our results
indicate that the
SFR does not peak at $z\sim 2$ but instead  continues to increase
 until $z \sim 10$.
We have used the burst formation rate to correct the observed GRB luminosity
function for
the incompleteness due to the detection threshold,   resulting in a 
luminosity function with a
power law index of $\sim -2.3$ that slightly rolls over at low
luminosities. The reality of our variability--luminosity relationship
requires confirmation
but, if valid, will provide a powerful tool for studying both GRBs and the
early universe.

\end{abstract}


\section{INTRODUCTION}

Thousands of gamma-ray bursts (GRBs) have been observed, but there is
no consensus on the physics underlying their defining 
characteristics:
huge luminosity, chaotic time history, and an energy release that
peaks in the gamma-ray energy range. 
The Italian--Dutch
BeppoSax satellite provided a crucial breakthrough in 1997 with the
discovery of precisely located X-ray afterglows (\cite{cos97}) and,
subsequently, afterglows across the spectrum from optical  
(\cite{vanpara97}) to radio (\cite{frail97}). See \cite{kulkarnihunt} for
a review. The optical afterglows led to measured
redshifts for a few afterglows and  
host
galaxies, firmly establishing that GRBs are cosmological in nature and the
largest explosions since the Big Bang. These few measured distances can be
used to calibrate other properties of GRBs that might serve as
standard candles. 

The peak flux, fluence, duration,
spectral hardness, and number of peaks in a GRB are clearly not standard
candles.
However, other properties of the time structure may be correlated
with
absolute luminosity. 
Based on an analysis of the time structure of BATSE bursts, \cite{sternn}
concluded that there was an intrinsic  correlation between luminosity and the
complexity of GRBs. 
 Norris et al.\ (1999, 2000) 
suggested that the
lag found from the cross-correlation of two spectral channels might be
correlated with luminosity. \cite{enricohuntsn} suggested that the
spikiness
of the time structure is correlated with luminosity, a relationship
similar to the Cepheid period--luminosity relationship, in which  
the  pulsation period of a star is correlated with luminosity. 
In our relationship, which is purely empirical, smooth bursts are
intrinsically less luminous.
We define  spikiness as 
effectively the
mean-square of the time signal after removing low frequencies by smoothing.

In this paper, we present the details of calculating  quantitative values 
for the
spikiness and its correlation with absolute luminosity. We use seven
Burst and Transient Source Experiment (BATSE) GRB events for which
redshifts 
are
known to calibrate the variability -- luminosity  relationship. We
apply
the relationship to the bright, long
GRBs in the BATSE catalog to obtain distances and luminosities. 
Based on only seven events, the
variability -- luminosity relationship  requires more
GRBs with redshifts to confirm its validity. We investigate 
some of
the implications of the relationship, in particular, the resulting
GRB luminosity function and the rate
of formation of GRB progenitors.

\section{THE VARIABILITY -- LUMINOSITY RELATIONSHIP}

We call the spikiness ``variability'' and calibrate it using GRBs with
known
redshifts ($z$) and  BATSE time histories. We must first correct the
observed time history
for cosmological  time dilation, correcting it to how it would appear at
some baseline redshift, $z_b$. 
The
most 
natural $z_b$ is, of course, zero. Other classes of objects, such as
supernovae, have most observed events at low redshifts, which  can be
used
to
calibrate  standard candles at $z_b = 0$. In the case of GRBs, all of
the events are
at quite high redshifts. The advantage of setting 
$z_b$ to, say, 2 is that smaller, and thus more stable, time
dilation corrections will be
applied. We define $Y$ to be
$(1+z)/(1+z_b)$ and we will
use $z_b = 2$ to correct the time histories for time dilation. (In previous 
versions of this paper we also scaled the luminosity to what it would 
appear to be from $z = z_b$. We no longer do that.)


The amount of stretching of the time history depends on $Y$ in two ways.
First, time
dilation causes a stretching equal to $Y$. Second, the spectrum is
redshifted such that the bandpass of BATSE records energies from $YE_L$ to
$YE_U$
rather than from $E_L$ to $E_U$, where 
$E_L$ to $E_U$ is the energy range of the BATSE time history. 
Bursts have a narrower time structure at higher energies (\cite{autofun}),
which affects the variability.
Let $V(E_L,E_U)$ be the variability  of the GRB time
structure in the bandpass $E_L$ to $E_U$. The energy-dependent correction
is $V(YE_L,YE_U)/V(E_L,E_U)$.  We can estimate
$V(YE_L,YE_U)/V(E_L,E_U)$ for $Y = 2$ because BATSE provides us with two
energy channels 
that happen to be shifted by a factor of 2 (i.e., channel 1 [25 to 50 keV]
and channel 2 [50 to 100 keV]). We selected 118 BATSE bursts that were
bright (peak flux greater than 1.5 photons s$^{-1}$ cm$^{-2}$), long
(90\% of the counts spread out over more than 20 s), with at least 100
points  5$\sigma$
above the background. We calculated the variability for both channel 1 and
channel
2 and determined that the median $V(YE_L,YE_U)/V(E_L,E_U)$ was $\sim
0.85$.
Furthermore, from symmetry, we
know that 
$V(Y^{-1}E_L,Y^{-1}E_U)/V(E_L,E_U) \sim V(E_L,E_U)/V(YE_L,YE_U)$, thus
channel 1 and channel 2 also gives us  the correction for $Y = 0.5$ (i.e.,
1/0.85).
We have little knowledge of $V(YE_L,YE_U)/V(E_L,E_U)$ for other values of
$Y$, but we do have the robust result that the average pulse width of a
GRB as a function
of energy is a power law (\cite{autofun}). Thus, we assume that the
correction to variability for the pulse spreading with energy is also
a power law:
%
\begin{equation}
V(YE_L,YE_U)/V(E_L,E_U) = Y^{-0.24}~~.
\label{KCORRECT}
\end{equation}
We have most confidence when $Y$ is less than 2 because we can use
channels 1 and 2 of BATSE to estimate the correction. If one uses $z_{b} =
0$, we could only confidently correct the energy dependency for events
from $z$ = 0 to $z$ = 1. Using $z_b = 2$ means that we can 
correct with high confidence the energy dependency for events
from $z$ = 0.5 to $z$ = 5.

To calculate the variability, we first 
fit a
linear or quadratic polynomial to the background in the non-burst portions
of the data. Let $g_i$ be
the gross counts observed in a 64 ms sample in the BATSE four channel data
(i.e., DISSC data, covering 25 to $\sim 800$ keV).  Let $b_i$ be the
counts
in a 64 ms background sample from the polynomial fit.
The net count is $c_i = g_i - b_i$. Using the known redshift, we correct for
time dilation by 
rebinning the counts  by stretching the time samples by
$Y$. Let $C_i$ be the new  stretched net count. Thus, $C_i$, represent what
the
time history would look like at $z=z_b$. Let $B_i$ be the stretched background.
The variability is then defined to be the (energy-corrected)
average
mean-square of the variations in $C_i$ relative to a smoothed time
history, that is,
%
\begin{equation}
V=Y^{-0.24}{1\over N}\sum{(C_i-<C>_{0.3T_{90}})^2 - (B_i+C_i) \over C^2_p}
\label{CTSTOV}
\end{equation}
where $C_p$ is the peak
of the (stretched)  net count during the burst, and
$<C>_{0.3T_{90}}$ is the count smoothed with a boxcar window with a length equal
to
30\% of the (stretched) $T_{90}$ duration of the burst. ($T_{90}$ is the
period that contains 90\% of the total counts.)
The $Y^{-0.24}$ term corrects the variability for the energy-dependence of
the time scale of a  GRB.
The $B_i+C_i$ term (stretched gross counts in a sample) accounts for
the Poisson noise (\cite{danvar}). The expected value of $V$
for pure noise is zero.
The sum is taken over the $N$
samples that exceed the background by at least 5$\sigma$. 

In Table 1, we list the BATSE GRBs with known redshifts and the
variability found using equation (\ref{CTSTOV}).

The associated luminosity of a GRB can be found from the redshift, the
observed spectral shape, and the observed peak photon flux averaged over
256 ms in a specified  energy range, $E_{L,P}$ to $E_{U,P}$
(i.e., $P_{256}$
photons s$^{-1}$ cm$^{-2}$ in the 50 to 300 keV band). The peak luminosity
(also
averaged over 256 ms and over the same energy range) per steradian is 
%
\begin{equation}
{L_{256}\over {\rm d}\Omega} = P_{256}<E>D^2  
\label{PTOLUM}
\end{equation}
where $D$ is the comoving distance and $<E>$ is the average photon energy
in the luminosity bandpass per photon in the count bandpass.
The comoving distance is
%
\begin{equation}
D = \int_0^z \big({c\over H_0}\big) { {\rm d}z \over 
  \sqrt{\Omega_{\Lambda} + \Omega_{\rm m}(1+z)^{3}} }~~,
\label{ZTODIS}
\end{equation}
in which we assume 
$\Omega_{\Lambda} = 0.7, \Omega_{\rm m} = 0.3$, and
$H_0$ = 65 km s$^{-1}$ Mpc$^{-1}$. From the
observed photon number spectrum, $\phi(E)$, 
%
\begin{equation}
<E> = { \int_{E_{L,P}}^{E_{U,P}} E \phi\big[{E\over 1+z}\big] {\rm d}E \over
        \int_{E_L}^{E_U} \phi[E] {\rm d}E }~~.
\label{AVEE}
\end{equation}
The observed spectra, $\phi[E]$, are characterized by the ``Band'' 
function
(\cite{band93}) defined by a low-energy spectral index, $\alpha$, a 
high-energy spectral index, $\beta$, and the peak of the $E^2\phi(E)$
distribution, $E_{\rm peak}$. 

In equation (\ref{PTOLUM}), 
we specify luminosity per steradian to emphasize that the angular extent
of the GRB emission is not known. An instrument  measures the flux
subtended
by a detector; that is, we always measure the luminosity per
solid angle. Thus, our variability -- luminosity relationship is
independent
of the (unknown) angular range that the GRB emits into. Often, the
equivalent
isotropic luminosity ($L_{4\pi}$) is quoted neglecting the effects of
beaming; 
$L_{4\pi} = 4\pi L/{\rm d}\Omega$. 
All our conclusions remain
valid if GRBs emit in a jet, but then our numerical results must be
multiplied by ${\rm d}\Omega_{\rm jet}/4\pi$, where ${\rm d}\Omega_{\rm
jet}$ is the jet opening angle.

In Table 1, we list $P_{256}, \alpha, \beta, E_{\rm peak}$, and the
resulting $L/{\rm d}\Omega$ and $V$ for each GRB. In Figure
\ref{burstzvarfun1},
we show the
variability as a
function
of $L/{\rm d}\Omega$ for the seven BATSE events with well known redshifts.
(An
eighth suggested redshift, for GRB980425, will be discussed later.)
The scatter in Figure \ref{burstzvarfun1} is much larger than we showed
in a preliminary report (\cite{enricohunts}). Some of the additional
scatter was introduced by  the addition of a new event (burst 7906). A
small amount is
due
to slight improvements in how we correct for the time dilation, but some
of
the differences are unexplained.

Through Monte Carlo simulations of these bursts we have determined that
the uncertainty in the position of the points due to counting statistics
is about the size of the plotting symbol.
If a relationship exists between absolute luminosity and variability,
then it
obviously has some scatter. In Figure \ref{burstzvarfun1} we have
suggested two power laws which probably bound the power law which best 
describes the real distribution. 
The two power laws are:
\begin{eqnarray}
\label{VTOLUMBOUND}
L/{\rm d}\Omega & = & 5.9 \times 10^{60}V^{5.8} {\rm erg~
s}^{-1}~~{\rm and}\\
                & = & 2.4 \times  10^{54}V^{2.2} {\rm erg~ s}^{-1}~~~~~.
\label{VTOLUMBOUNDN}
\end{eqnarray}
These curves are only to guide the eye. \cite{danvarn} has done a statistical
analysis of the seven events used here plus
some additional upper limits to find a true confidence region for the 
variability --
luminosity relationship. More events will be
required to accurately define the real distribution.

Figure \ref{burstzvarfun2} shows the seven bursts used above and 
BATSE burst 6707 (GRB980425), which has been associated with the
nearby
supernova, SN 1998bw. This association is controversial
(\cite{galama980425sn,woosley980425,kulkarni98,graziani2000}).
If
the association is correct, then GRB980425 was much less luminous than
other GRBs (see discussion in \cite{bloom98}).
However, its variability is also much smaller, following our general
trend that low variability implies low absolute luminosity. We feel that
GRB980425 adds believability to our relationship in proportion to the
probability
that the association with SN 1998bw is true.

\section{Comparison to Different Definition of Variability}

Recently, \cite{danvarn} has proposed an alternative definition of variability.
Both this paper and \cite{danvarn}  relate the variability to the
square of the time history
after removing low frequencies by smoothing. There are three major areas where the
definitions differ; (1)the time scale for smoothing, (2) the normalization to make
the definition roughly independent of distance, and (3) how time dilation and
intrinsic peak-spread with energy are accommodated.  \cite{danvarn} also does
precise statistics at every step and determines confidence regions for all
results. The \cite{danvarn} definition can be heuristically expressed as
%
\begin{equation}
V_R = {1 \over N_E} \sum_{\Delta E_i = 1}^{\Delta E_i = N_E} 
{ \sum\big[<c_{\Delta E_i}>_{(1+z)^{\beta}} - <c_{\Delta E_i}>_{T_{f=0.45}}\big]^2
\over \sum \big[<c_{\Delta E_i}>_{(1+z)^{\beta}}\big]^2  }
\label{DANVAR}
\end{equation}
where $N_E$ is the number of energy channels,
$c_{\Delta E_i}$ is the counts in the $i$-th energy bin, and $<c_{\Delta
E_i}>_{(1+z)^{\beta}}$ indicates that the counts have been smoothed by a box car
function with a width of $(1+z)^{\beta}$.
This formulation is only for the purpose of
indicating the major differences in the definitions and the original formulation
should be consulted for details. For example, this formulation 
omits the various terms
associated with the propagation of Poisson noise and simplifies others.

The first difference is the time scale for smoothing. We use 30\% of the $T_{90}$
duration provided by the BATSE catalog. \cite{danvarn} defines a duration 
(``$T_{f=0.45}$'')
equal to
the
smallest fraction of the burst time history that contains a fraction $f$ of the
total counts. If all bursts had time histories that monotonically
increased and
then decreased,
then $T_{90}$ would be equal to  $T_{f=0.90}$. The
\cite{danvarn}
time scale  reflects the total on-time that the burst is active whereas
$T_{90}$ reflects the total range of time that the burst is active. Bursts with
precursors might have a large $T_{90}$ but a small $T_{f=0.90}$. \cite{danvarn}
finds that $f = 0.45$ gives a robust definition of variability. It is too early to
tell if this different way of setting the smoothing time scale is important. 

The second difference is in the normalization in the denominator. We use the
square of the peak counts ($C_p^2$) whereas \cite{danvarn} uses the sum of the
counts
squared. Note, that the \cite{danvarn} normalization is not the square of the
fluence (which
would be $(\sum c_i)^2$). 
We also define the variability to be an average variation per sample (note the
$1/N$ term in eq.~[\ref{CTSTOV}]). Thus, the difference in normalization is
roughly
$NC_p^2/\sum C_i^2$. Since $N$ will scale as the duration of the
event, our variability values will be smaller.
Again, it is too early to
tell if this difference in normalization is important.

The key conceptual difference is how the time dilation and the peak spreading with
energy is handled. This paper treats these issues separately. We use the known $z$
to rebin
the observed time history to remove the time dilation (i.e., the conversion of the
observed counts, $c_i$, to those corrected for time dilation, $C_i$). We then
remove the effects of the peak spreading by the $Y^{-0.24}$ term in equation
(\ref{CTSTOV}). That term was derived based on how our definition
of variability varied with energy in bright BATSE bursts. We mitigate these
effects by correcting the time history to how it would appear at an intermediate
$z_b$ rather than $z = 0$.  \cite{danvarn} applies a smoothing to the time history
to account for the  combined effects of time dilation  and peak spreading.
Time dilation uniformly stretches the whole time history by $(1+z)^1$ and peak
spreading
stretches the individual peaks by $\sim(1+z)^{-0.4}$ (\cite{autofun}).
Thus, in equation (\ref{DANVAR}), the time history is
smoothed by $(1+z)^{\beta}$ where $\beta = 0.6$.
With such smoothing, \cite{danvarn} finds that each energy range gives about the
same variability, so the variability from multiple energy ranges are
combined with a Bayesian inference formalism which is heuristically 
similar to a
weighted average (represented by  ${1\over N_E}\sum$ in
eq.~[\ref{DANVAR}]).

There are other minor differences. In the definition of luminosity, we use
$P_{256}$ whereas \cite{danvarn} uses $P_{1024}$. Also, we use a source
frame energy range of 50 to 300 keV and the measured Band parameters
whereas \cite{danvarn} uses 100 to 1000
keV and an average spectral shape.

Figure ~\ref{comparedan} is a scatter plot of our values of variability {\it vs.}
the \cite{danvarn} values. Although there are differences in how
variability is
defined, the two methods give highly correlated values. Apparently all three
differences in the definition are required to have such a correlation. If the
\cite{danvarn} time scale and normalization are used in equation
(\ref{CTSTOV}),
the
values are much less correlated.

\section{DETERMINING DISTANCE FROM GRB TIME STRUCTURE}

The variability -- luminosity relationship allows us to determine the
distance to GRBs based only on the observed time history and parameters
determined from the gamma-ray observations ($P_{256}, T_{90}, \alpha,
\beta,
E_{\rm peak}$). 

The first step in determining distance from the time structure is to 
assume 
that $z = z_b$ and to
calculate the variability, V, using equation (\ref{CTSTOV}). 
For the variability -- luminosity relationship, we have used a power law
lying intermediate between the two power laws shown in Figure
\ref{burstzvarfun1}:
\begin{equation}
{L_{4\pi} \over 4\pi} = 
L/{\rm d}\Omega  =  3.1 \times 10^{56}V^{3.35} {\rm erg~ s}^{-1}~.
\label{VTOLUM}
\end{equation}
Here, the luminosity is the peak erg s$^{-1}$ per steradian in a
specified (source frame) energy bandpass
($E_{L,P}$ = 50 keV to $E_{U,P}$ = 300 keV), averaged over 256 ms.
The  peak photon flux is
%
\begin{equation}
P_{256} = { 3.1 \times 10^{56}V^{3.35} \over <E> D^2}
~~{\rm photon~ cm}^{-2}~ {\rm s}^{-1}~~.
\label{LUMTOP}
\end{equation}
We substitute equations (\ref{CTSTOV}, \ref{ZTODIS}, \ref{AVEE})
into equation (\ref{LUMTOP}), to solve for $z$, the only unknown.
Once we obtain $z$ (i.e., $Y$), we iterate using  equation
(\ref{CTSTOV}) to
compute an improved estimate of the variability. Equation (\ref{LUMTOP})
is solved again
and the iterative process is repeated until $z$ converges. In 19 cases, the
solution converged to a value of $z$ larger than 12. These cases were assigned a
value of $z$ equal to 12 in the table and in the figures. In our
analysis 
to obtain the formation rate and the luminosity function,  we do
not use any part of the parameter space beyond $z = 11.5$.

\section{REDSHIFTS AND LUMINOSITIES FOR 220 BATSE GRBs}

The redshifts determined  from GRBs events detected by the 
BeppoSax satellite were only obtained for long bursts, so our
luminosity -- variability  
relationship may only be valid  for long bursts. Also, variability is
more difficult to estimate for short bursts. Thus, we  estimate 
redshifts only for long ($T_{90} > 20$ s),  bright ($P_{256} > 1.5$
photons s$^{-1}$ cm$^{-2}$) BATSE bursts. There are 227 such bursts in the
BATSE 4B catalog (\cite{fourthcat}).
Four  did not have at least 10 samples with a $5\sigma$
significance (bursts 1145, 1546, 1626, 5572), two bursts produced negative
variability (bursts 2106 and 2863), and the iterations to solve equation
(\ref{LUMTOP}) failed for one burst (2476).
These events
were not considered further. 
We did not have  ``Band''
parameters available, so we used the default values
$\alpha=1.5,
\beta=2.0$, and
$E_{\rm peak} = 250$ keV.

Table 2 lists the properties of the 220 GRBs and
Figure \ref{lumz} shows the resulting distribution of luminosities and
redshifts as open squares.
The solid squares are the seven GRBs (omitting GRB980425) from Table 1.
The luminosities are smaller than often quoted for GRBs because we have
evaluated the luminosity assuming a finite energy range
(50
to 300 keV) rather than bolometric. The solid lines are lines of constant
$P_{256}$ starting at $P_{256} = 1.5$ photon cm$^{-2}$ s$^{-1}$. 
If we did not have a detector threshold, then the projection of the
GRBs onto the luminosity axis would directly give the GRB luminosity
function. Although it is possible to imagine other scenarios, we
expect the  luminosity function to be the same for all redshifts.
The projection of Figure \ref{lumz} onto the redshift axis (divided by
the volume in the redshift range 0 to $z$), would give the GRB formation
rate which is analogous to
the star formation rate, SFR (see \cite{madau96}). We expect every range
of luminosity
to
have
a similar GRB formation rate. The BATSE detector threshold is
obviously
preventing us from
seeing many events at high $z$, as well as preventing the direct
determination
of the 
luminosity function and the burst formation rate.

Previously, we did not have estimated redshifts for so many GRBs, and so
we
were forced to only analyze the brightness distribution (i.e., \lognp).
All we knew was the number of GRBs between the solid lines (i.e., in
ranges of $P_{256}$) but not their distribution in the $L_{4\pi}-z$ space.
Hence, previous studies (e.g.,
\cite{totani97,wijerssfr,hogg,lamb99}) could only check for
consistency between the
observed distribution of $P_{256}$ by assuming both a GRB
luminosity function (usually a power law) {\it and}  that the
burst
formation rate followed an assumed SFR.
Indeed, we have analyzed the \lognp\ distribution from 
the 220 GRBs based on the solid
lines in Figure \ref{lumz} and found that it gives similar results to analyses
using a larger number of  BATSE GRBs: the best fit power law index of the
luminosity function is -2.4
and the dynamic range of the power law was a factor of 170.

The $L_{4\pi}-z$ distribution in Figure \ref{lumz}  allow us to
directly determine the GRB luminosity function and the burst formation
without  assuming 
that the GRB formation rate follows the SFR or assuming a form for the
luminosity function.

\subsection{Consistency with the Star Formation rate}

The luminosity -- redshift distribution in Figure \ref{lumz} is complete
at
a particular redshift down to a threshold luminosity, $L_{th}(z)$,
which follows the $P_{256}$ = 1.5 photons cm$^{-2}$ s$^{-1}$
line in Figure \ref{lumz}. 
We restrict
ourselves to BATSE events brighter than 1.5 photons s$^{-1}$ cm$^{-2}$, 
and so we
are confident that BATSE is complete to the corresponding $L_{256}$.
Figure \ref{lumzbound} shows the 220 BATSE GRBs divided into ranges of $z$
and  $L$. Consider the dotted lines: they divide the bursts into
ranges of $z$ each selected to contain approximately the same number of
events. There are seven ranges of $z$ bound by $z_1, z_2, ..., z_8$ =
0.3, 0.75, 1.25, 2.0, 3.0, 5.0, 7.0, 10.0. Within the $i$-th range
(bounded by $z_i$ to $z_{i+1}$), the figure is complete to 
$L = L_{th}(z_{i+1})$. Within each
range we can find
the \lognl\  distribution for $L_{256} > L_{th}$
without making any corrections or assumptions. We normalize
each distribution by the rate-volume between $z_i$ and $z_{i+1}$, that is,
we divide by
%
\begin{equation}
N_{z_i,z_{i+1}} = 
\int_{z_i}^{z_{i+1}} {4\pi D^2 \rho(z) {\rm d}z
    \over (1+z)\sqrt{\Omega_{\Lambda} + \Omega_{\rm m}(1+z)^{3}} }
\label{VOLSFR}
\end{equation}
where $\rho(z)$ is the comoving density of events.
The $1+z$ term in the denominator accounts for the fact
that we observe an event rate.

In Figure \ref{bfrvolt}, we show the resulting 
\lognl\ distributions if the co-moving density, $\rho(z)$, is constant.
The
true
\lognl\ distribution is, plausibly, independent
of $z$, 
whereas the distributions in Figure \ref{bfrvolt} are clearly inconsistent
with each other. Of course, the likely explanation is that the density of
GRBs is not constant. In Figure \ref{bfr} we show the resulting
\lognl\  distributions if $\rho(z)$ follows the SFR, $\rho_{\rm SFR}(z)$,
for which we have used 
the observations summarized by
\cite{steidel} as parameterized by \cite{rowanrobin}.
 (Other fits to the observations including estimates of the SFR in the
sub-mm
waveband, such as
\cite{blain99}, give
similar results.)
The resulting distributions are more consistent than in Figure
\ref{bfrvolt},
but still not consistent with each other. Clearly, bursts in $z$-ranges \# 
4, 5, 6, and 7 (i.e., $z > 2$) are occurring at a relatively higher rate than
$z$-ranges \# 1, 2,
and 3 (i.e., $z < 2$).

In previous work, it was necessary to simultaneously assume that
GRBs follow the SFR  {\it and} a luminosity function, usually 
a bounded, power law. In that case,  one could only check for
consistency with the number of events found within bands such as the solid
lines in Figure \ref{lumz} (i.e., the \lognp\ distribution, see
\cite{totani97,wijerssfr,hogg,lamb99}). Non-evolving standard candle
models would also fit
(\cite{fenlognp,fenbloom}), although the occurrence of high redshift GRBs
indicated that non-evolving models were probably not acceptable
(\cite{martin}).
Figure \ref{bfrvolt} demonstrates that a non-evolving,
standard
candle model is inconsistent with the observations. Figure \ref{bfr} demonstrates
for the first time that GRBs could be occurring at a higher rate than star
formation at $z > 2$.

The normalization constants that bring the distributions of 
Figure \ref{bfrvolt} into agreement directly give 
the relative burst formation
rate as a function of $z$. Figure \ref{bfrburstz} demonstrates this
process. In  Figure \ref{bfrburstz}, we have shifted the curves of  Figure
\ref{bfrvolt} until they all fall on a common curve. The resulting factors
are plotted in   Figure \ref{bfrsfr}  as the solid squares.
The solid squares roughly follow $(1+z)^{3.3 \pm 0.3}$ and is the burst
formation rate derived solely from gamma-ray data without assuming a
star formation rate or a form for the luminosity function.
The solid histogram uses equation (\ref{VTOLUM}). The dotted histograms
use equations (\ref{VTOLUMBOUND}) and (\ref{VTOLUMBOUNDN}), so represent
the systematic uncertainity in the burst formation rate from the
uncertainity in the variability -- luminosity relationship.
The dashed line is the  SFR from \cite{blain99} that is consistent with
current sub-mm observations  and the solid
line is from \cite{steidel}.
 The dotted curve in Figure \ref{bfrsfr} is a theoretical estimate of the
star formation rate at high $z$ (from \cite{gnedin}).
(We have arbitrarily normalized the rates to
each other at $z = 1$.)

For $z > 2$,
the burst formation rate is much higher than the SFR and continues to rise well
beyond where other observations indicate that star formation has decreased or
become constant.  However, recent results indicate that the SFR may,
indeed, rise until large $z$. The  SFR may be underestimated at
high redshifts ($z > 3$) due to an under correction in previous surveys
of cosmological surface brightness dimming. When corrected, the resulting
SFR increases to $\sim z = 10$, similar to our
result (Lanzetta, K., private communication). 

  If the SFR follows the burst formation rate, many more stars would have
formed earlier in the universe. One can integrate the SFR based on our
burst formation rate (i.e., the histogram in Fig. ~[\ref{bfrburstz}]) from
$z =0 $ to $\sim\infty$ to determine $\Omega_{*}$, the
density parameter of stars (cf. \cite{madau96}). We find that our burst
formation rate give
$\Omega_{*} = 0.0062 \pm 0.0005$ whereas the density contained in the
stellar population based on the observed luminous mass is believed to be
$\sim  0.0045 \pm 0.0014$
(\cite{fukugita98,trentham00}).
Current Models used to obtain the SFR (such as \cite{steidel}) give values of
$\Omega_{*}$ between
0.0028 and 0.0035. Thus, current SFR predictions give values that are about 
$1 \sigma$ lower than those obtained  from the luminous mass estimates, and our
burst formation
rate gives a value that is about $1 \sigma$ higher.
Our burst formation rate probably reflects the formation of massive stars
that become black holes whereas the SFR mostly reflect less massive
stars. Thus, these two formation rates sample different parts of the mass
function of stars.

GRBs are likely to provide a more robust estimate for the SFR at
redshifts where the optical estimates of the SFR are affected by dust
(see, for example, detailed calculations by \cite{lamb99,blain99}).
However, the relationship might be complex because there could be an
offset time between the star formation and the GRB phase (especially if
the mechanism is neutron star -- black hole collisions) and also might
depend on
the metallicity (see \cite{bloom99}).

\subsection{GRB Luminosity Function}

The solid lines in Figure \ref{lumzbound} divide the bursts into ranges
of luminosity. There are nine ranges, spaced logarithmically. Within the
$i$-th range (bounded by $L_i$ to $L_{i+1}$), the figure is complete to
$z_c$ defined by  
$L_{i} = L_{th}(z_c)$. The largest $z_c$ is 11.5. If there were no threshold, we
could count the relative number of events in each range of luminosity and
determine the GRB luminosity function directly. Thus, to estimate the GRB
luminosity function, we must correct for the events that are missing due
to the detector 
threshold. 
For each range of
luminosity, we find the number of GRBs between $z = 0$
and $z_c$. We
then correct that number by the relative amount of star formation for
$z<z_c$ compared to the total amount of star formation to large $z$. Using
equation (\ref{VOLSFR}), we multiply the number of 
GRBs observed at $z < z_c$ by $N_{0,\infty}/N_{0,z_c}$.
The normalized number of GRBs was divided by the exposure time of the 4B
BATSE catalog
(2.56 yr) and by the width of the luminosity bin in units of $10^{50}$ erg
s$^{-1}$.

The
results are shown in Figure \ref{lumfun}. 
The solid line uses the SFR
of \cite{steidel} in equation (\ref{VOLSFR}). The
dotted line uses the GRB formation rate from Figure \ref{bfrsfr}.
Previously, studies had to assume a form for the luminosity function
(e.g., a bounded power law) and assume that the GRB formation rate
follows the SFR. Here, we have determined the GRB luminosity function
without assuming a form for it or assuming that GRB follow the SFR. The
dotted line is derived entirely from gamma-ray data.

Both the luminosity function based on the SFR and the luminosity function
entirely from gamma-ray data are similar.
Above $L = 4 \times 10^{51}$ erg
s$^{-1}$, the luminosity function is roughly a power law with an index of 
about -1.9 when using the SFR and an index
of about  -2.3 when using the GRB formation rate.
At low
luminosity, the observed distribution falls somewhat below the
extensions of the power laws.
Previous studies 
indicated that the luminosity function must be bounded
(see e.g, \cite{loredo98,martin,lamb99}).
For example, \cite{martinn} argued that the observations of redshifts as high as
3.4 (GRB991214) requires density evolution and a broad luminosity function. 
Only models in \cite{martinn} that used a luminosity function similar to our
Figure \ref{lumfun} (i.e., a range of 3 orders of magnitude and a high luminosity
slope of 2) gave reasonable probabilities for observing an event with a redshift
of 3.4. Since our burst formation rate increases beyond $z =1$ while the SFR of
\cite{martinn} remained constant for $z > 1$, we would expect even higher
estimates for the probability of observing events out to $z = 3.4$.
In Figure \ref{lumfun} we
might be seeing the gradual roll over of the distribution at low
luminosity. 

\section{TIME DILATION?}

\cite{norris94n} binned BATSE bursts by $P_{256}$ and searched for the 
effects of time
dilation. If GRBs are standard candles, then there should be a one-to-one
relationship between $P_{256}$ and $z$. Bursts with smaller values of
$P_{256}$ should appear  longer due to the time dilation
caused by the expansion of the universe. Indeed, \cite{norris94n} found
significant time dilation by effectively  binning GRBs in
a
manner similar to the solid lines in Figure \ref{lumz}. Our
results clearly indicate that GRBs are not standard candles. By binning
the events
by $z$ (as in Figure \ref{lumzbound}), we should be able to make a
direct
segregation of events by $z$ and, therefore, see time dilation that is not
smeared out by having events from different $z$ in each sample.

We  use the aligned peak test (\cite{igor93}) to characterize the time
scale in each $z$
bin. In the aligned peak test, one shifts the time histories until the
largest peaks are aligned and then an average is taken of the time
histories. Figure \ref{alignpeakb} shows the aligned peaks for the first six
ranges of $z$ from Figure \ref{lumzbound}. (The seventh bin has fewer
events.) All $z$-ranges give about the same temporal structure. 
 The
time structure is affected by the expansion of the universe in two ways.
First, the expansion introduces a time dilation stretching of $1+z$.
Second, the
time histories are redshifted by a factor of $1+z$ so that events at low
$z$ appear to be destretched  because there is a tendency for
time histories to have wider peaks at lower energy. \cite{autofunn} found
that the redshift effect on the aligned peak width scales as
$(1+z)^{-0.42}$. Thus, one would 
expect the aligned peak width to be stretched by $(1+z)^{0.58}$. In Figure
\ref{alignpeakz}, we destretched each burst by a factor of 
$(1+z)^{-0.58}$ before
aligning and averaging.

Since Figure \ref{alignpeakz} has been corrected for time dilation, we would
expect it to have a tighter clustering of curves.
Comparing Figures \ref{alignpeakb} and \ref{alignpeakz}, it appears
that correcting for the amount of time stretching that we expect
actually makes the averaged aligned peaks  more inconsistent. Thus,
this test does not support that variability provides valid redshifts.
This test ought to be fairly sensitive because of the wide range of $z$
that we cover. 
The typical redshift
in the first range  is
0.5 and the typical redshift in the sixth is 6, so one
should expect nearly a factor of $(7/1.5)^{0.58} =2.4$ difference in
the pulse widths.
The test might be weak  for several reasons. First,
usually
one
needs at least 40
GRBs in each sample to average out the burst-to-burst variations. In our
six samples, there are only 16, 21, 22, 25, 30, and 16 GRBs.
Second, the pulse width could be an intrinsic
property of the bursts that depends strongly on luminosity, and thus, less
weakly on $z$ (\cite{lambrome}).
Perhaps with more bursts 
and using a
lower threshold, we would be able to  make a stronger test.

\section{SUMMARY}

The  variability -- luminosity relationship could be a
powerful tool for understanding GRBs.
We have presented a method for characterizing the variability of a burst
(Eq.~[\ref{CTSTOV}]) and related it to the absolute GRB luminosities.
(Fig.~\ref{burstzvarfun1}). 
The variability -- luminosity relationship (eq. \ref{VTOLUM}) is based on
only seven events, so its real test will come from future GRBs with known
redshifts. In the interim, one can check if it produces reasonable results
for the GRB luminosity function and the GRB
formation rate. 
We have presented the details for converting the observables (i.e., 
$T_{90},P_{256}, \alpha, \beta, E_{\rm peak}$, and the observed net
counts) to the GRB 
distance  using equations (\ref{CTSTOV}) and (\ref{LUMTOP}). The
resulting
luminosities and redshifts for 220 bright, long BATSE
GRBs are shown in Table 2 and  Figure \ref{lumz}. 

The luminosity -- redshift distribution allows us to 
find the brightness distribution
(\lognl) in several ranges of $z$. 
If we assume no evolution, then the distributions are
clearly inconsistent with each other (Fig.\ \ref{bfrvolt}). 
Assuming reasonable SFRs  
improves the consistency as a function of $z$, but apparently
the SFR underestimates the true rate of GRBs at $z > 2$ (Fig.\
\ref{bfr}). 
Unlike previous work, the GRB
formation rate is estimated without making any assumptions 
about either the  luminosity function or the SFR. 
The resulting GRB formation rate (Fig.\
\ref{bfrsfr}) is roughly a power law, $\sim (1+z)^{3.3 \pm 0.3}$.
It continues to rise at redshifts 
where various optical observations indicate that the SFR levels off.
Unlike the
SFR inferred from optical observations, the GRB formation rate is not 
affected by dust. 

Based on a power law luminosity function and the SFR,
\cite{kommersn} concluded from the Log~$N$--Log~$P$~ distribution
that BATSE was
detecting most (50\% to 70\%) of the GRBs that occur. Our burst formation
rate implies that BATSE is seeing only a small fraction of the GRBs.
Assuming a BATSE threshold of $P_{256} = 0.5$ photons s$^{-1}$ cm$^{-2}$
and conservatively using a power law luminosity function
(index = -1.9)
that extends from
$10^{50}$ to $10^{52.6}$ erg s$^{-1}$, we find that BATSE is seeing only
18\% of the bursts with $z < 2$ and only 3\% of bursts with $z < 10$. 
An instrument with a threshold 5 times smaller, would see 58\% of the
bursts with $z < 2$ and  11\% of bursts with $z < 10$.
Thus, such an
instrument (e.g., Swift, see \cite{neilswift})
would see about 3  to 4 times more bursts per steradian than BATSE.
With a luminosity function power law with an index of -2.3, such an
instrument would see 5 times more bursts than BATSE.

We have used the GRB formation rate and the SFR to correct the observed
GRB luminosity function for
the incompleteness due to the detector threshold. The resulting luminosity
function
(Fig.\ \ref{lumfun})
is a power law with an index of $\sim -2.3$ above $4 \times 10^{51}$ erg
s$^{-1}$ and rolls over slightly at lower luminosities.

The reality of our variability -- luminosity relationship is far from
certain.
It is based on only seven events, although GRB980425 (if associated with SN
1998bw) follows the same trend. We failed to detect the expected time
dilation implied by
our variability -- luminosity relationship (Figs.\ \ref{alignpeakb} and
\ref{alignpeakz}), but that test is weaken by the few
number of bursts that define each aligned peak average. More GRBs 
redshifts are needed to test whether
variability or correlation lags (\cite{norris99}) can, indeed, give
redshifts to GRBs solely from  gamma-ray observations.

The underlying cause of a luminosity -- variability relationship is
unclear.
In the context of the internal shock model, larger initial Lorentz factors
tend to produce more efficient collisions and, for the same number of
initial shells, the time histories are more variable.
(\cite{ksp97}).  We have run such models, including deceleration
(c.f., \cite{fenintext}), and
find that simply varying the Lorentz factors, ambient density,
and/or the initial mass of the shells does not produce the dynamic range
of variability that is observed. Apparently, the observed 
luminosity -- variability  
correlation is set deep in the center engine.

{\it Acknowledgments}: We recognize useful conversations and insights
from  A. Blain, J. Bloom, D. Lamb, J. Norris, A. Merloni, M. Rees, D.
Reichart,
P. Madau, and P. Natarajan. We thank D. Lamb for suggesting the term
``Cepheid-like relationship''. We thank D. Reichart for pointing out the
importance of subtracting the contribution from Poisson noise from the
variability measure. D. Reichart and Brad Schafer pointed out serious problems in
the
original manuscript that caused many of the $z$ values to be smaller.

\clearpage
\begin{table}
\def\tablerule{\noalign{\hrule}}
\centerline{Table 1}
\centerline{Burst Parameters}
{\vskip 0.1truecm}
 \label{tablereds}
 \begin{tabular}{@{}lccccccccc}
\tablerule
\noalign{\vskip 0.01truecm}
\tablerule
\noalign{\vskip 0.1truecm}
\tablerule
\noalign{\vskip 0.01truecm}
\tablerule
  &BATSE& &&&&&& & \\
 GRB &Number&$z$&$T_{90}$&$P_{256}^1$&$\alpha^2$&$\beta^2$&$E_{peak}^2$
&Variability&$L/d\Omega$\\
\tablerule
\tablerule
970508$^3$&6225&0.835&13.3&1.2&-1.18&-1.88&137.2&0.0114&
$5.75\times10^{49}$\\
970828$^4$&6350&0.958&8.76&4.9&-0.87&-2.08&176.0&0.0173&
$2.95\times10^{50}$\\
971214$^5$&6533&3.412&29.9&2.3&-1.14&-2.85&106.8&0.0195&
$2.81\times10^{51}$\\
980425$^6$&6707&0.0085&20.6&1.1&-1.50&-2.00&250.0&0.0036&
$2.63\times10^{45}$\\
980703$^7$&6891&0.967&51.2&2.6&-1.32&-2.04&181.0&0.0168&
$1.82\times10^{50}$\\
990123$^8$&7343&1.600&63.7&16.4&-1.06&-2.30&267.7&0.0276&
$2.88\times10^{51}$\\
990510$^9$&7560&1.62&60.6&8.2&-1.43&-2.49&74.2&0.0608&
$2.57\times10^{51}$\\
991216$^{10}$&7906&1.02&40.3&85.1&-1.50&-2.00&250.0&0.0268&
$6.76\times10^{51}$\\
\tablerule
\tablerule
 \end{tabular}

 \medskip
$^1$ Peak photon flux (ph cm$^{-2}$ s$^{-1}$) on 256 ms time scale, 50
- 300 keV \vfill\break
$^2$ Time resolved fits from 128 channel spectral data (R. Preece, private
communication)\vfill\break
$^3$ \cite{metzger97} \vfill\break
$^4$ \cite{odewahn97} \vfill\break
$^5$ \cite{kulkarni98} \vfill\break
$^7$ \cite{djorgovski} \vfill\break
$^6$ \cite{galama980425}, default Band parameters \vfill\break
$^8$ \cite{kulkarni0123} \vfill\break
$^9$ \cite{vreewijka} \vfill\break
$^{10}$ \cite{vreewijkb}
\end{table}

\clearpage
 \clearpage
 \begin{table}
 \def\tablerule{\noalign{\hrule}}

 \centerline{TABLE 2}
 \centerline{Redshifts and Luminosities of BATSE  GRBs from Variability$^{1}$}
 {\vskip 0.1truecm}
 \begin{tabular}{@{}lccccclccccc}
 \tablerule
 \noalign{\vskip 0.01truecm}
 \tablerule
 \noalign{\vskip 0.1truecm}
 \tablerule
 \noalign{\vskip 0.01truecm}
 \tablerule
 \# &$P_{256}$ & $T_{90}$ & $V$ & $z$ & $L_{4\pi}$& \# & $P_{256}$ & $T_{90}$ & $V$ & $z$ & $L_{4\pi}$ \\
 \tablerule
 \noalign{\vskip 0.01truecm}
 \tablerule
  109&    3.62&    90.2&    0.0233&    1.8&$ 1.35 \times 10^{52}$ &
  130&    3.47&    62.0&    0.0168&    1.1&$ 4.52 \times 10^{51}$\\
 \noalign{\vskip -0.001truecm}
  143&   47.57&    50.8&    0.0598&    2.3&$ 3.16 \times 10^{53}$ &
  219&   18.06&    29.7&    0.0172&    0.6&$ 4.84 \times 10^{51}$\\
 \noalign{\vskip -0.001truecm}
  222&    3.99&    73.1&    0.0380&    3.6&$ 6.92 \times 10^{52}$ &
  249&   34.62&    28.5&    0.0170&    0.4&$ 4.67 \times 10^{51}$\\
 \noalign{\vskip -0.001truecm}
  394&    4.78&   106.1&    0.0172&    1.0&$ 4.91 \times 10^{51}$ &
  398&    1.71&    25.6&    0.0175&    1.6&$ 5.18 \times 10^{51}$\\
 \noalign{\vskip -0.001truecm}
  467&    7.73&    37.7&    0.0312&    2.0&$ 3.59 \times 10^{52}$ &
  503&    5.05&    46.7&    0.0753&   10.2&$ 6.85 \times 10^{53}$\\
 \noalign{\vskip -0.001truecm}
  548&    2.00&    36.2&    0.0162&    1.4&$ 3.99 \times 10^{51}$ &
  563&    1.89&    22.5&    0.0115&    0.9&$ 1.25 \times 10^{51}$\\
 \noalign{\vskip -0.001truecm}
  647&    7.07&    59.6&    0.0293&    1.9&$ 2.92 \times 10^{52}$ &
  660&    4.55&    20.2&    0.0305&    2.4&$ 3.31 \times 10^{52}$\\
 \noalign{\vskip -0.001truecm}
  676&    4.20&    77.8&    0.0254&    1.9&$ 1.81 \times 10^{52}$ &
  678&    6.18&    53.9&    0.0226&    1.4&$ 1.21 \times 10^{52}$\\
 \noalign{\vskip -0.001truecm}
  761&    3.21&    81.5&    0.0364&    3.8&$ 6.02 \times 10^{52}$ &
  869&    3.52&   110.5&    0.0557&    7.3&$ 2.50 \times 10^{53}$\\
 \noalign{\vskip -0.001truecm}
  907&    3.57&   158.1&    0.0386&    3.9&$ 7.34 \times 10^{52}$ &
  973&    5.29&    90.0&    0.0462&    4.3&$ 1.33 \times 10^{53}$\\
 \noalign{\vskip -0.001truecm}
 1141&    9.01&    20.1&    0.0142&    0.6&$ 2.55 \times 10^{51}$ &
 1150&    1.71&    61.6&    0.0366&    5.1&$ 6.10 \times 10^{52}$\\
 \noalign{\vskip -0.001truecm}
 1157&   10.04&   170.6&    0.0697&    6.3&$ 5.31 \times 10^{53}$ &
 1235&    2.53&   190.0&    0.0854&   12.0&$ 1.04 \times 10^{54}$\\
 \noalign{\vskip -0.001truecm}
 1288&    6.55&   318.6&    0.0269&    1.7&$ 2.18 \times 10^{52}$ &
 1385&    3.62&    51.8&    0.0192&    1.4&$ 7.04 \times 10^{51}$\\
 \noalign{\vskip -0.001truecm}
 1396&    1.68&    21.0&    0.0418&    6.4&$ 9.54 \times 10^{52}$ &
 1419&    4.51&   122.5&    0.0642&    8.2&$ 4.02 \times 10^{53}$\\
 \noalign{\vskip -0.001truecm}
 1440&   11.50&    24.4&    0.0526&    3.7&$ 2.06 \times 10^{53}$ &
 1447&    1.74&    22.7&    0.0151&    1.3&$ 3.16 \times 10^{51}$\\
 \noalign{\vskip -0.001truecm}
 1467&    2.26&    26.8&    0.0176&    1.5&$ 5.26 \times 10^{51}$ &
 1468&    3.34&    51.1&    0.0519&    6.6&$ 1.97 \times 10^{53}$\\
 \noalign{\vskip -0.001truecm}
 1533&    4.00&    99.8&    0.0489&    5.5&$ 1.62 \times 10^{53}$ &
 1541&   35.58&    26.2&    0.0266&    0.8&$ 2.10 \times 10^{52}$\\
 \noalign{\vskip -0.001truecm}
 1578&    3.75&    20.5&    0.0131&    0.8&$ 1.96 \times 10^{51}$ &
 1579&    3.25&    73.5&    0.0614&    9.0&$ 3.46 \times 10^{53}$\\
 \noalign{\vskip -0.001truecm}
 1601&    2.14&    48.5&    0.0542&    9.0&$ 2.28 \times 10^{53}$ &
 1606&    7.82&    88.6&    0.0200&    1.0&$ 8.08 \times 10^{51}$\\
 \noalign{\vskip -0.001truecm}
 1623&    2.98&    66.1&    0.0291&    2.7&$ 2.85 \times 10^{52}$ &
 1652&    4.08&    67.7&    0.0244&    1.8&$ 1.57 \times 10^{52}$\\
 \noalign{\vskip -0.001truecm}
 1660&    1.67&    53.9&    0.0669&   12.0&$ 4.61 \times 10^{53}$ &
 1663&   19.00&    36.0&    0.0249&    1.0&$ 1.69 \times 10^{52}$\\
 \noalign{\vskip -0.001truecm}
 1676&   10.49&    52.8&    0.0386&    2.4&$ 7.30 \times 10^{52}$ &
 1712&    3.10&   307.6&    0.0341&    3.4&$ 4.83 \times 10^{52}$\\
 \noalign{\vskip -0.001truecm}
 1733&    3.00&    36.9&    0.0392&    4.4&$ 7.70 \times 10^{52}$ &
 1734&    1.70&    46.7&    0.0876&   12.0&$ 1.14 \times 10^{54}$\\
 \noalign{\vskip -0.001truecm}
 1819&    2.01&    53.0&    0.1347&   12.0&$ 4.81 \times 10^{54}$ &
 1886&   16.37&   275.7&    0.1072&   10.3&$ 2.24 \times 10^{54}$\\
 \noalign{\vskip -0.001truecm}
 1967&    1.72&    25.4&    0.0146&    1.3&$ 2.80 \times 10^{51}$ &
 1982&    1.68&   109.8&    0.0279&    3.3&$ 2.45 \times 10^{52}$\\
 \noalign{\vskip -0.001truecm}
 1989&    2.73&   272.4&    0.0534&    7.7&$ 2.17 \times 10^{53}$ &
 1993&    1.69&    30.5&    0.0371&    5.3&$ 6.39 \times 10^{52}$\\
 \noalign{\vskip -0.001truecm}
 1997&   16.35&    60.2&    0.0195&    0.7&$ 7.46 \times 10^{51}$ &
 2047&    2.12&    41.1&    0.0703&   12.0&$ 5.45 \times 10^{53}$\\
 \noalign{\vskip -0.001truecm}
 2061&    2.19&   174.7&    0.0185&    1.6&$ 6.23 \times 10^{51}$ &
 2067&   18.10&    30.9&    0.0132&    0.4&$ 2.00 \times 10^{51}$\\
 \noalign{\vskip -0.001truecm}
 2080&    5.64&    53.8&    0.0186&    1.1&$ 6.39 \times 10^{51}$ &
 2090&   10.15&    38.1&    0.0607&    4.9&$ 3.33 \times 10^{53}$\\
 \noalign{\vskip -0.001truecm}
 2122&    1.89&   120.6&    0.0187&    1.7&$ 6.42 \times 10^{51}$ &
 2123&    2.12&    22.0&    0.0032&    0.1&$ 0.18 \times 10^{50}$\\
 \noalign{\vskip -0.001truecm}
 2138&    7.00&    77.3&    0.0313&    2.1&$ 3.61 \times 10^{52}$ &
 2156&   16.57&   154.4&    0.0282&    1.2&$ 2.55 \times 10^{52}$\\
 \noalign{\vskip -0.001truecm}
 2190&    1.82&   135.9&    0.0262&    2.9&$ 2.00 \times 10^{52}$ &
 2193&    1.55&   135.2&    0.0219&    2.4&$ 1.10 \times 10^{52}$\\
 \noalign{\vskip -0.001truecm}
 2213&    4.59&    61.9&    0.0457&    4.5&$ 1.29 \times 10^{53}$ &
 2228&    8.10&    90.1&    0.0262&    1.5&$ 1.99 \times 10^{52}$\\
 \noalign{\vskip -0.001truecm}
 2232&    6.02&    56.5&    0.0651&    7.2&$ 4.21 \times 10^{53}$ &
 2287&    1.91&   462.0&    0.0337&    4.2&$ 4.65 \times 10^{52}$\\
 \noalign{\vskip -0.001truecm}
 2316&    3.83&    29.2&    0.0024&    0.1&$ 0.64 \times 10^{49}$ &
 2328&    1.57&    54.0&    0.0230&    2.6&$ 1.30 \times 10^{52}$\\
 \noalign{\vskip -0.001truecm}
 2329&   40.40&    22.1&    0.0183&    0.4&$ 5.96 \times 10^{51}$ &
 2340&    1.61&    22.5&    0.0490&    8.8&$ 1.63 \times 10^{53}$\\
 \noalign{\vskip -0.001truecm}
 2345&    2.49&    89.0&    0.0953&   12.0&$ 1.51 \times 10^{54}$ &
 2346&    2.93&    83.1&    0.0504&    6.7&$ 1.78 \times 10^{53}$\\
 \end{tabular}
 \end{table}
 \clearpage
 \begin{table}
 \def\tablerule{\noalign{\hrule}}

 \centerline{TABLE 2 {\it continued}}
 \centerline{Redshifts and Luminosities of BATSE  GRBs from Variability$^{1}$}
 {\vskip 0.1truecm}
 \begin{tabular}{@{}lccccclccccc}
 \tablerule
 \noalign{\vskip 0.01truecm}
 \tablerule
 \noalign{\vskip 0.1truecm}
 \tablerule
 \noalign{\vskip 0.01truecm}
 \tablerule
 \# &$P_{256}$ & $T_{90}$ & $V$ & $z$ & $L_{4\pi}$& \# & $P_{256}$ & $T_{90}$ & $V$ & $z$ & $L_{4\pi}$ \\
 \tablerule
 \noalign{\vskip 0.01truecm}
 \tablerule
 2362&    3.35&    45.3&    0.0640&    9.5&$ 3.97 \times 10^{53}$ &
 2383&    3.06&    50.9&    0.0333&    3.3&$ 4.48 \times 10^{52}$\\
 \noalign{\vskip -0.001truecm}
 2387&    3.86&    42.0&    0.0098&    0.5&$ 7.51 \times 10^{50}$ &
 2428&    2.05&   144.9&    0.0617&   11.6&$ 3.51 \times 10^{53}$\\
 \noalign{\vskip -0.001truecm}
 2436&    6.08&    33.3&    0.0221&    1.3&$ 1.13 \times 10^{52}$ &
 2443&    2.10&    28.9&    0.0245&    2.5&$ 1.60 \times 10^{52}$\\
 \noalign{\vskip -0.001truecm}
 2450&    7.57&    45.9&    0.0262&    1.5&$ 1.99 \times 10^{52}$ &
 2451&    2.82&    35.0&    0.0487&    6.5&$ 1.59 \times 10^{53}$\\
 \noalign{\vskip -0.001truecm}
 2511&    1.79&    64.3&    0.0714&   12.0&$ 5.74 \times 10^{53}$ &
 2519&    1.53&    90.2&    0.0595&   12.0&$ 3.11 \times 10^{53}$\\
 \noalign{\vskip -0.001truecm}
 2522&    3.23&    83.1&    0.0272&    2.4&$ 2.27 \times 10^{52}$ &
 2530&    1.86&   194.9&    0.0323&    4.0&$ 4.04 \times 10^{52}$\\
 \noalign{\vskip -0.001truecm}
 2533&    8.92&    74.3&    0.0133&    0.5&$ 2.06 \times 10^{51}$ &
 2581&    1.66&   104.4&    0.0504&    9.0&$ 1.79 \times 10^{53}$\\
 \noalign{\vskip -0.001truecm}
 2593&    1.56&    29.9&    0.0376&    5.6&$ 6.72 \times 10^{52}$ &
 2603&    3.19&    24.7&    0.0082&    0.4&$ 4.01 \times 10^{50}$\\
 \noalign{\vskip -0.001truecm}
 2606&    2.38&   134.7&    0.0193&    1.6&$ 7.15 \times 10^{51}$ &
 2665&    1.99&    30.6&    0.0398&    5.5&$ 8.10 \times 10^{52}$\\
 \noalign{\vskip -0.001truecm}
 2681&    1.64&    55.5&    0.0427&    6.8&$ 1.03 \times 10^{53}$ &
 2700&    4.06&    57.9&    0.0371&    3.4&$ 6.38 \times 10^{52}$\\
 \noalign{\vskip -0.001truecm}
 2703&    2.89&    57.7&    0.0215&    1.8&$ 1.02 \times 10^{52}$ &
 2780&    1.59&    53.6&    0.0179&    1.7&$ 5.57 \times 10^{51}$\\
 \noalign{\vskip -0.001truecm}
 2790&    3.47&    45.3&    0.0333&    3.1&$ 4.46 \times 10^{52}$ &
 2798&   23.75&    49.2&    0.0230&    0.8&$ 1.29 \times 10^{52}$\\
 \noalign{\vskip -0.001truecm}
 2812&   10.52&    30.7&    0.0364&    2.2&$ 5.99 \times 10^{52}$ &
 2831&   43.43&   150.1&    0.0249&    0.7&$ 1.69 \times 10^{52}$\\
 \noalign{\vskip -0.001truecm}
 2855&    9.53&    42.5&    0.0175&    0.8&$ 5.19 \times 10^{51}$ &
 2877&    2.92&   114.1&    0.0356&    3.8&$ 5.56 \times 10^{52}$\\
 \noalign{\vskip -0.001truecm}
 2889&    5.92&    75.9&    0.0125&    0.6&$ 1.68 \times 10^{51}$ &
 2890&    2.32&    51.6&    0.0224&    2.1&$ 1.19 \times 10^{52}$\\
 \noalign{\vskip -0.001truecm}
 2891&    8.71&    60.7&    0.0314&    1.9&$ 3.65 \times 10^{52}$ &
 2897&    2.94&    28.0&    0.0178&    1.3&$ 5.44 \times 10^{51}$\\
 \noalign{\vskip -0.001truecm}
 2913&    5.20&    22.9&    0.0449&    4.2&$ 1.22 \times 10^{53}$ &
 2922&    2.85&   160.8&    0.0399&    4.6&$ 8.19 \times 10^{52}$\\
 \noalign{\vskip -0.001truecm}
 2929&    5.91&    42.8&    0.0185&    1.0&$ 6.25 \times 10^{51}$ &
 2958&    3.75&    36.9&    0.0406&    4.2&$ 8.69 \times 10^{52}$\\
 \noalign{\vskip -0.001truecm}
 2984&    4.61&    32.8&    0.0302&    2.4&$ 3.21 \times 10^{52}$ &
 2993&    3.22&    44.8&    0.0249&    2.1&$ 1.68 \times 10^{52}$\\
 \noalign{\vskip -0.001truecm}
 2994&   14.42&    48.6&    0.0245&    1.1&$ 1.59 \times 10^{52}$ &
 3001&    4.19&    29.7&    0.0337&    2.9&$ 4.63 \times 10^{52}$\\
 \noalign{\vskip -0.001truecm}
 3003&    2.83&    37.6&    0.0108&    0.7&$ 1.02 \times 10^{51}$ &
 3011&    1.68&    49.7&    0.0433&    6.9&$ 1.08 \times 10^{53}$\\
 \noalign{\vskip -0.001truecm}
 3015&    1.75&    26.8&    0.0453&    7.3&$ 1.25 \times 10^{53}$ &
 3035&    6.03&    88.4&    0.0198&    1.1&$ 7.82 \times 10^{51}$\\
 \noalign{\vskip -0.001truecm}
 3040&    1.70&    26.0&    0.0376&    5.4&$ 6.69 \times 10^{52}$ &
 3042&    6.74&   182.0&    0.0417&    3.3&$ 9.50 \times 10^{52}$\\
 \noalign{\vskip -0.001truecm}
 3055&    1.78&    40.6&    0.0047&    0.2&$ 0.61 \times 10^{50}$ &
 3056&    2.41&    36.3&    0.0196&    1.7&$ 7.58 \times 10^{51}$\\
 \noalign{\vskip -0.001truecm}
 3057&   32.36&    34.9&    0.0256&    0.8&$ 1.84 \times 10^{52}$ &
 3067&   18.67&    67.0&    0.0451&    2.3&$ 1.23 \times 10^{53}$\\
 \noalign{\vskip -0.001truecm}
 3075&    2.32&    43.6&    0.0304&    3.3&$ 3.30 \times 10^{52}$ &
 3093&    2.03&    89.3&    0.0431&    6.2&$ 1.06 \times 10^{53}$\\
 \noalign{\vskip -0.001truecm}
 3101&    2.22&    37.4&    0.0661&   12.0&$ 4.43 \times 10^{53}$ &
 3115&   11.10&    45.3&    0.0557&    4.1&$ 2.50 \times 10^{53}$\\
 \noalign{\vskip -0.001truecm}
 3128&   12.41&    32.2&    0.0365&    2.0&$ 6.07 \times 10^{52}$ &
 3142&    2.03&    33.1&    0.0424&    5.9&$10.00 \times 10^{52}$\\
 \noalign{\vskip -0.001truecm}
 3178&   14.34&    39.9&    0.0225&    0.9&$ 1.20 \times 10^{52}$ &
 3212&    2.02&    83.4&    0.0416&    5.9&$ 9.38 \times 10^{52}$\\
 \noalign{\vskip -0.001truecm}
 3227&   17.03&   115.3&    0.0260&    1.1&$ 1.95 \times 10^{52}$ &
 3237&    2.01&    46.2&    0.0582&   10.3&$ 2.89 \times 10^{53}$\\
 \noalign{\vskip -0.001truecm}
 3241&   12.48&    45.4&    0.0368&    2.0&$ 6.25 \times 10^{52}$ &
 3245&   12.79&    77.1&    0.0207&    0.9&$ 9.00 \times 10^{51}$\\
 \noalign{\vskip -0.001truecm}
 3255&   11.87&    34.9&    0.0248&    1.2&$ 1.66 \times 10^{52}$ &
 3256&    1.76&    96.6&    0.0965&   12.0&$ 1.58 \times 10^{54}$\\
 \noalign{\vskip -0.001truecm}
 3257&    3.06&    63.6&    0.0290&    2.7&$ 2.81 \times 10^{52}$ &
 3283&    2.57&    45.8&    0.0633&   10.8&$ 3.83 \times 10^{53}$\\
 \noalign{\vskip -0.001truecm}
 3287&    6.69&    33.4&    0.0200&    1.1&$ 8.13 \times 10^{51}$ &
 3290&   10.70&    43.7&    0.0803&    7.6&$ 8.51 \times 10^{53}$\\
 \noalign{\vskip -0.001truecm}
 3301&    2.81&    36.1&    0.0268&    2.5&$ 2.17 \times 10^{52}$ &
 3306&    3.28&   108.5&    0.0279&    2.5&$ 2.47 \times 10^{52}$\\
 \noalign{\vskip -0.001truecm}
 3307&    1.70&    43.0&    0.0572&   11.2&$ 2.74 \times 10^{53}$ &
 3319&    2.11&   206.3&    0.2210&   12.0&$ 2.53 \times 10^{55}$\\
 \noalign{\vskip -0.001truecm}
 3330&    6.75&    62.0&    0.0196&    1.1&$ 7.58 \times 10^{51}$ &
 3345&    6.76&    40.4&    0.0359&    2.6&$ 5.72 \times 10^{52}$\\
 \end{tabular}
 \end{table}
 \clearpage
 \begin{table}
 \def\tablerule{\noalign{\hrule}}

 \centerline{TABLE 2 {\it continued}}
 \centerline{Redshifts and Luminosities of BATSE  GRBs from Variability$^{1}$}
 {\vskip 0.1truecm}
 \begin{tabular}{@{}lccccclccccc}
 \tablerule
 \noalign{\vskip 0.01truecm}
 \tablerule
 \noalign{\vskip 0.1truecm}
 \tablerule
 \noalign{\vskip 0.01truecm}
 \tablerule
 \# &$P_{256}$ & $T_{90}$ & $V$ & $z$ & $L_{4\pi}$& \# & $P_{256}$ & $T_{90}$ & $V$ & $z$ & $L_{4\pi}$ \\
 \tablerule
 \noalign{\vskip 0.01truecm}
 \tablerule
 3351&    4.48&   102.9&    0.0354&    3.1&$ 5.50 \times 10^{52}$ &
 3352&    3.71&    46.3&    0.0047&    0.2&$ 0.62 \times 10^{50}$\\
 \noalign{\vskip -0.001truecm}
 3405&    1.53&    67.4&    0.0622&   12.0&$ 3.61 \times 10^{53}$ &
 3407&    1.53&    91.4&    0.0435&    7.3&$ 1.09 \times 10^{53}$\\
 \noalign{\vskip -0.001truecm}
 3408&   12.73&    58.6&    0.0283&    1.4&$ 2.59 \times 10^{52}$ &
 3415&    9.16&    54.3&    0.0353&    2.2&$ 5.42 \times 10^{52}$\\
 \noalign{\vskip -0.001truecm}
 3436&    3.56&    40.0&    0.0264&    2.2&$ 2.06 \times 10^{52}$ &
 3439&    1.66&   150.8&    0.0933&   12.0&$ 1.41 \times 10^{54}$\\
 \noalign{\vskip -0.001truecm}
 3448&    2.19&   331.2&    0.0313&    3.5&$ 3.64 \times 10^{52}$ &
 3458&    8.68&   673.8&    0.0709&    6.9&$ 5.62 \times 10^{53}$\\
 \noalign{\vskip -0.001truecm}
 3481&   21.94&    40.8&    0.0397&    1.8&$ 8.03 \times 10^{52}$ &
 3488&    8.65&    60.2&    0.0521&    4.2&$ 2.00 \times 10^{53}$\\
 \noalign{\vskip -0.001truecm}
 3489&    6.65&    33.1&    0.0147&    0.7&$ 2.89 \times 10^{51}$ &
 3512&    4.84&    29.8&    0.0417&    3.8&$ 9.51 \times 10^{52}$\\
 \noalign{\vskip -0.001truecm}
 3523&   21.57&    59.1&    0.0208&    0.7&$ 9.21 \times 10^{51}$ &
 3569&    4.53&    21.0&    0.0832&   12.0&$ 9.59 \times 10^{53}$\\
 \noalign{\vskip -0.001truecm}
 3593&    6.61&    59.1&    0.0450&    3.7&$ 1.23 \times 10^{53}$ &
 3618&    2.50&    90.2&    0.0335&    3.7&$ 4.56 \times 10^{52}$\\
 \noalign{\vskip -0.001truecm}
 3634&    3.30&   142.0&    0.0518&    6.6&$ 1.96 \times 10^{53}$ &
 3648&    5.70&    57.1&    0.0297&    2.1&$ 3.04 \times 10^{52}$\\
 \noalign{\vskip -0.001truecm}
 3649&    4.28&    26.5&    0.0297&    2.4&$ 3.02 \times 10^{52}$ &
 3662&    3.05&    68.9&    0.0503&    6.5&$ 1.77 \times 10^{53}$\\
 \noalign{\vskip -0.001truecm}
 3663&    4.48&   204.4&    0.0452&    4.5&$ 1.24 \times 10^{53}$ &
 3664&    1.98&    85.2&    0.0452&    6.8&$ 1.24 \times 10^{53}$\\
 \noalign{\vskip -0.001truecm}
 3745&    2.21&    84.3&    0.0430&    5.9&$ 1.05 \times 10^{53}$ &
 3765&   25.29&    72.6&    0.0357&    1.4&$ 5.62 \times 10^{52}$\\
 \noalign{\vskip -0.001truecm}
 3788&    5.20&    65.7&    0.0187&    1.1&$ 6.44 \times 10^{51}$ &
 3843&    2.33&    50.5&    0.0179&    1.5&$ 5.54 \times 10^{51}$\\
 \noalign{\vskip -0.001truecm}
 3853&    3.08&    91.3&    0.1850&   12.0&$ 1.39 \times 10^{55}$ &
 3860&    4.44&    31.5&    0.0168&    1.0&$ 4.47 \times 10^{51}$\\
 \noalign{\vskip -0.001truecm}
 3879&    1.51&    30.9&    0.0407&    6.5&$ 8.70 \times 10^{52}$ &
 3891&   13.69&    41.5&    0.0305&    1.5&$ 3.32 \times 10^{52}$\\
 \noalign{\vskip -0.001truecm}
 3893&    3.70&    35.7&    0.0109&    0.6&$ 1.05 \times 10^{51}$ &
 3900&    1.53&    86.7&    0.0682&   12.0&$ 4.93 \times 10^{53}$\\
 \noalign{\vskip -0.001truecm}
 3905&    4.39&    24.3&    0.0359&    3.1&$ 5.71 \times 10^{52}$ &
 3906&    2.32&    20.7&    0.0218&    2.0&$ 1.07 \times 10^{52}$\\
 \noalign{\vskip -0.001truecm}
 3912&    4.04&    26.0&    0.0376&    3.5&$ 6.69 \times 10^{52}$ &
 3918&    2.00&   110.5&    0.0249&    2.6&$ 1.67 \times 10^{52}$\\
 \noalign{\vskip -0.001truecm}
 3929&    3.97&    28.5&    0.0197&    1.3&$ 7.71 \times 10^{51}$ &
 3954&    8.19&    29.7&    0.0472&    3.6&$ 1.44 \times 10^{53}$\\
 \noalign{\vskip -0.001truecm}
 4039&    5.45&    58.8&    0.0278&    1.9&$ 2.45 \times 10^{52}$ &
 4157&    2.27&    20.0&    0.0130&    0.9&$ 1.91 \times 10^{51}$\\
 \noalign{\vskip -0.001truecm}
 4216&    1.51&    25.5&    0.0347&    5.0&$ 5.14 \times 10^{52}$ &
 4350&    3.27&    52.0&    0.0418&    4.6&$ 9.55 \times 10^{52}$\\
 \noalign{\vskip -0.001truecm}
 4368&   56.23&    36.5&    0.0183&    0.4&$ 6.04 \times 10^{51}$ &
 4701&    7.14&    74.7&    0.0400&    3.0&$ 8.26 \times 10^{52}$\\
 \noalign{\vskip -0.001truecm}
 5304&   13.28&    22.8&    0.0166&    0.6&$ 4.34 \times 10^{51}$ &
 5389&    4.14&    23.7&    0.0406&    4.0&$ 8.68 \times 10^{52}$\\
 \noalign{\vskip -0.001truecm}
 5419&    3.43&    22.8&    0.0211&    1.6&$ 9.65 \times 10^{51}$ &
 5421&    1.97&   266.8&    0.0536&    9.2&$ 2.20 \times 10^{53}$\\
 \noalign{\vskip -0.001truecm}
 5433&    2.83&    76.0&    0.1560&   12.0&$ 7.87 \times 10^{54}$ &
 5447&    3.88&    55.9&    0.0299&    2.5&$ 3.10 \times 10^{52}$\\
 \noalign{\vskip -0.001truecm}
 5450&    4.15&   172.2&    0.0712&   10.3&$ 5.69 \times 10^{53}$ &
 5451&    4.30&    24.5&    0.0226&    1.6&$ 1.21 \times 10^{52}$\\
 \noalign{\vskip -0.001truecm}
 5464&    2.60&    28.9&    0.0382&    4.5&$ 7.07 \times 10^{52}$ &
 5470&    4.79&    30.4&    0.0677&    8.7&$ 4.80 \times 10^{53}$\\
 \noalign{\vskip -0.001truecm}
 5472&    1.96&    55.1&    0.0371&    4.9&$ 6.43 \times 10^{52}$ &
 5473&    3.79&    80.6&    0.0554&    6.9&$ 2.46 \times 10^{53}$\\
 \noalign{\vskip -0.001truecm}
 5475&    2.42&    72.8&    0.0552&    8.7&$ 2.43 \times 10^{53}$ &
 5476&    2.55&    21.9&    0.0331&    3.6&$ 4.39 \times 10^{52}$\\
 \noalign{\vskip -0.001truecm}
 5478&    2.96&   297.0&    0.0688&   11.6&$ 5.08 \times 10^{53}$ &
 5479&    2.76&   137.3&    0.0476&    6.3&$ 1.47 \times 10^{53}$\\
 \noalign{\vskip -0.001truecm}
 5484&    2.68&    64.6&    0.0656&   11.2&$ 4.32 \times 10^{53}$ &
 5486&    9.35&    86.7&    0.0354&    2.2&$ 5.48 \times 10^{52}$\\
 \noalign{\vskip -0.001truecm}
 5489&    9.44&   140.5&    0.0470&    3.4&$ 1.41 \times 10^{53}$ &
 5495&    2.12&    26.3&    0.0701&   12.0&$ 5.39 \times 10^{53}$\\
 \noalign{\vskip -0.001truecm}
 5518&    2.35&    51.9&    0.0531&    8.3&$ 2.13 \times 10^{53}$ &
 5526&    3.38&    72.4&    0.0281&    2.4&$ 2.54 \times 10^{52}$\\
 \noalign{\vskip -0.001truecm}
 5531&    1.67&    33.9&    0.0124&    1.0&$ 1.62 \times 10^{51}$ &
 5539&    1.88&    77.8&    0.1066&   12.0&$ 2.20 \times 10^{54}$\\
 \noalign{\vskip -0.001truecm}
 5541&    1.66&    26.4&    0.0349&    4.8&$ 5.24 \times 10^{52}$ &
 5548&    5.95&   187.8&    0.0384&    3.0&$ 7.16 \times 10^{52}$\\
 \tablerule
 \noalign{\vskip 0.01truecm}
 \tablerule
 \end{tabular}
 $^{1}$ Electronic version of table available from efenimore@lanl.gov \hfill
 \end{table}

%

\clearpage
\figcaption[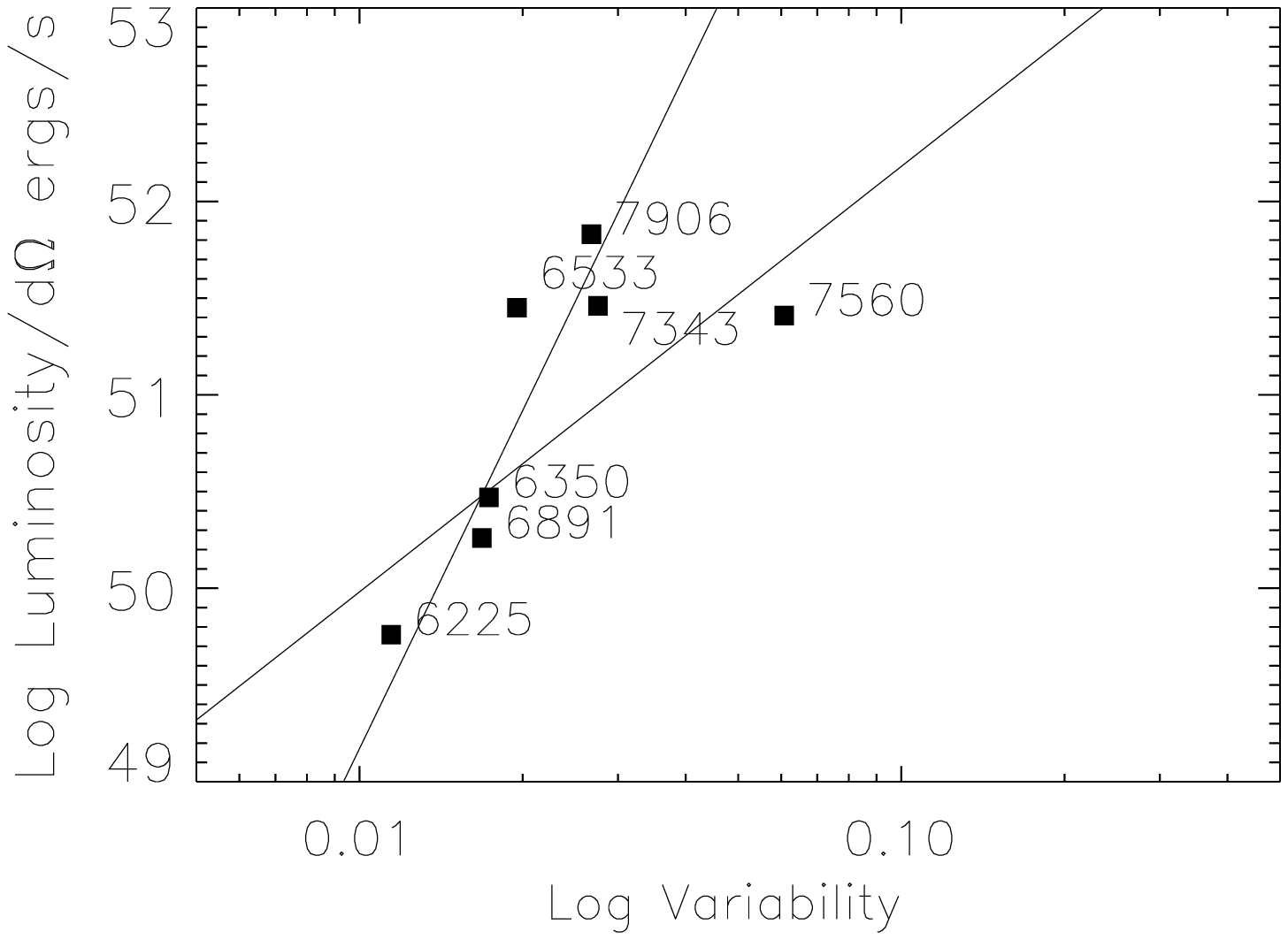]{
Luminosity and variability for BATSE bursts with known redshifts. The six
bursts are labeled by the BATSE trigger numbers (see Table 1). 
The two solid lines bound the region that
contains the average of the distribution. Many more bursts will be
necessary to define the average of this distribution more accurately.
\label{burstzvarfun1}
}

\figcaption[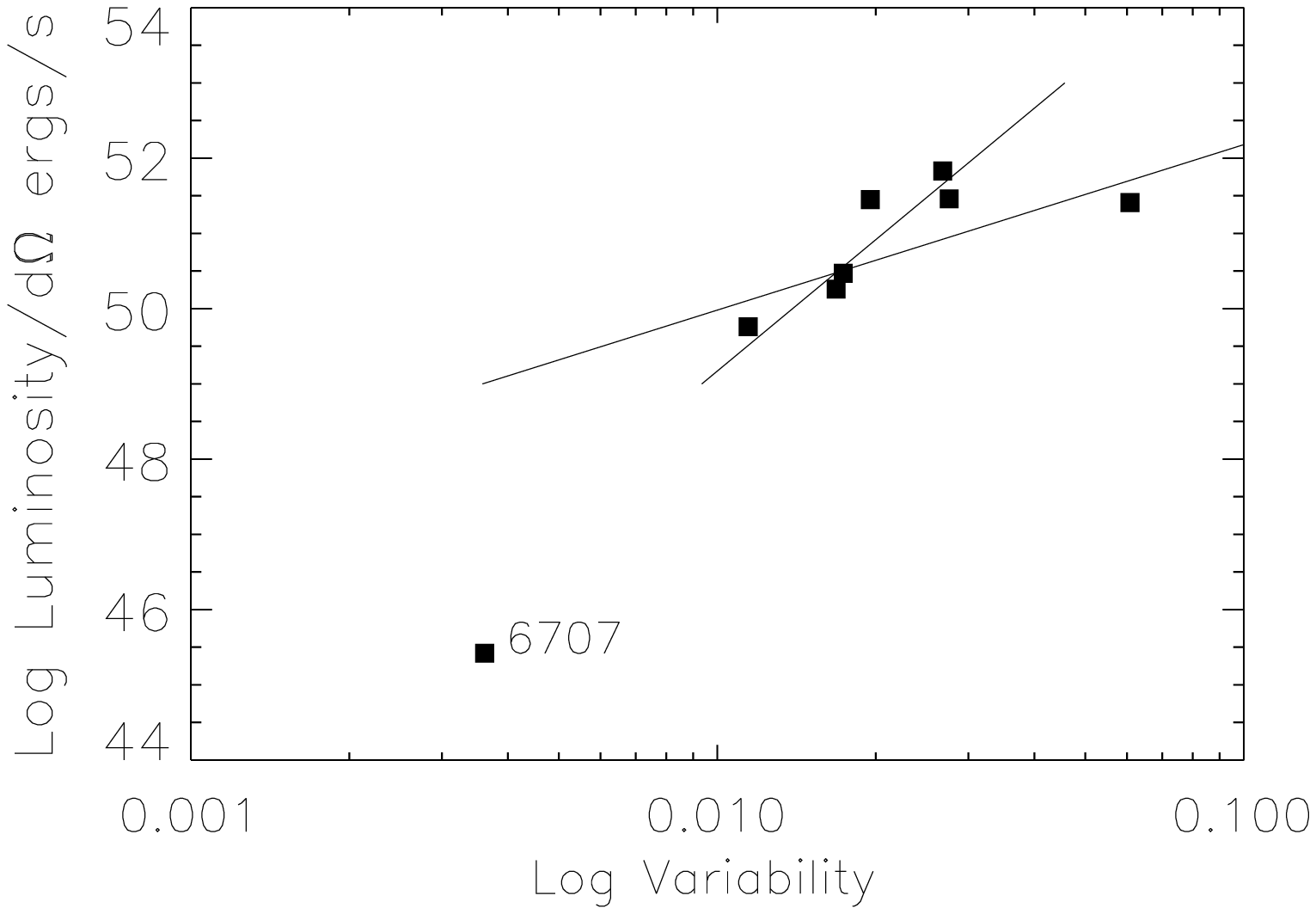]{
Luminosity and variability for BATSE bursts including GRB980425 (BATSE
trigger number 6707). If GRB980425 is associated with SN 1998bw, it is
very close ($z = 0.0085$) and much less luminous than other GRBs. The
association is controversial
(\cite{galama980425sn,woosley980425,kulkarni98,graziani2000}).
GRB980425  has the lowest observed variability, which is in general
agreement with our proposed trend.
\label{burstzvarfun2}
}

\figcaption[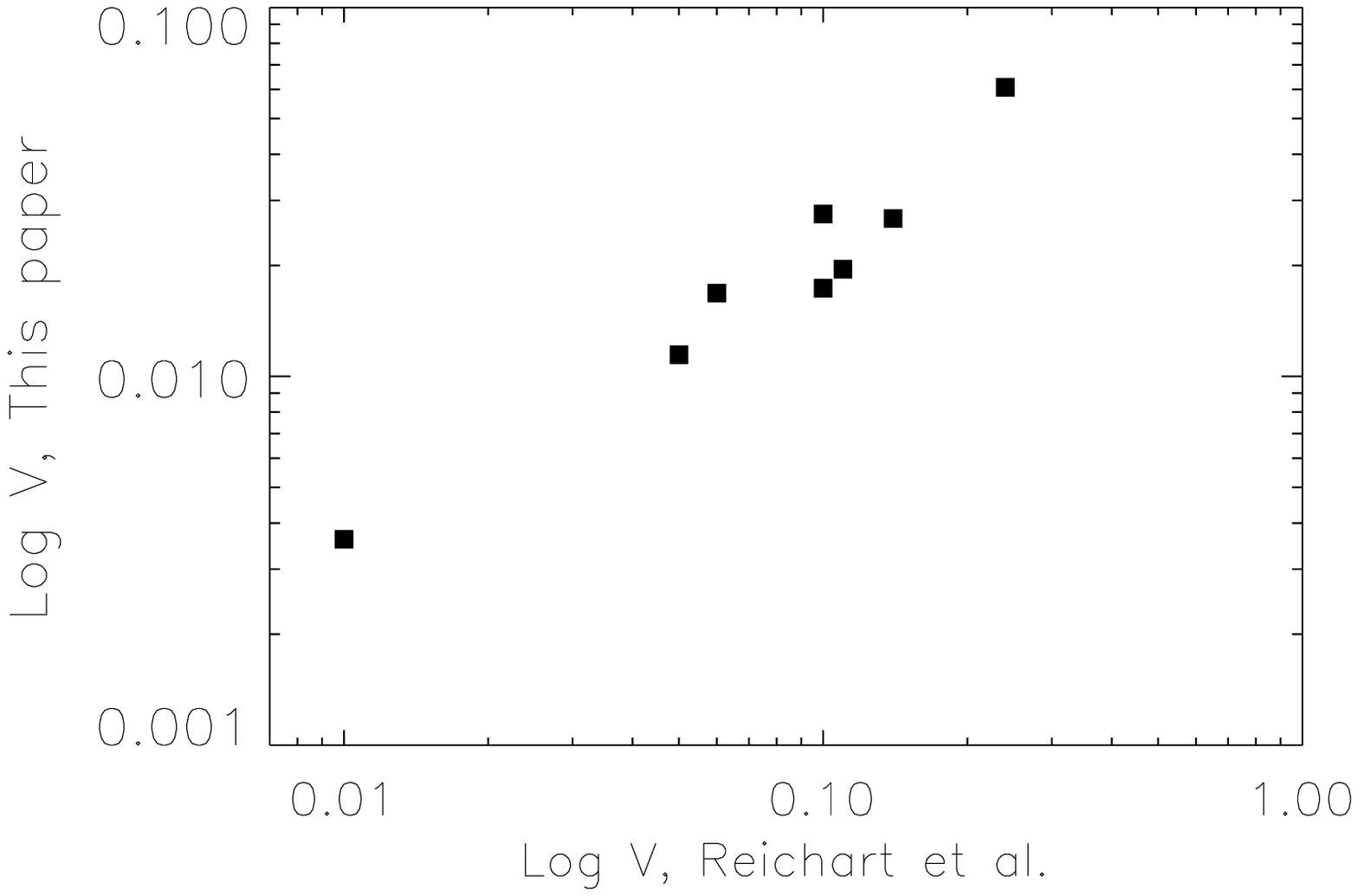]{
Comparison of variability from this paper and that of \cite{danvar}.
Although defined somewhat differently, the two definitions are highly
correlated.
\label{comparedan} }

\figcaption[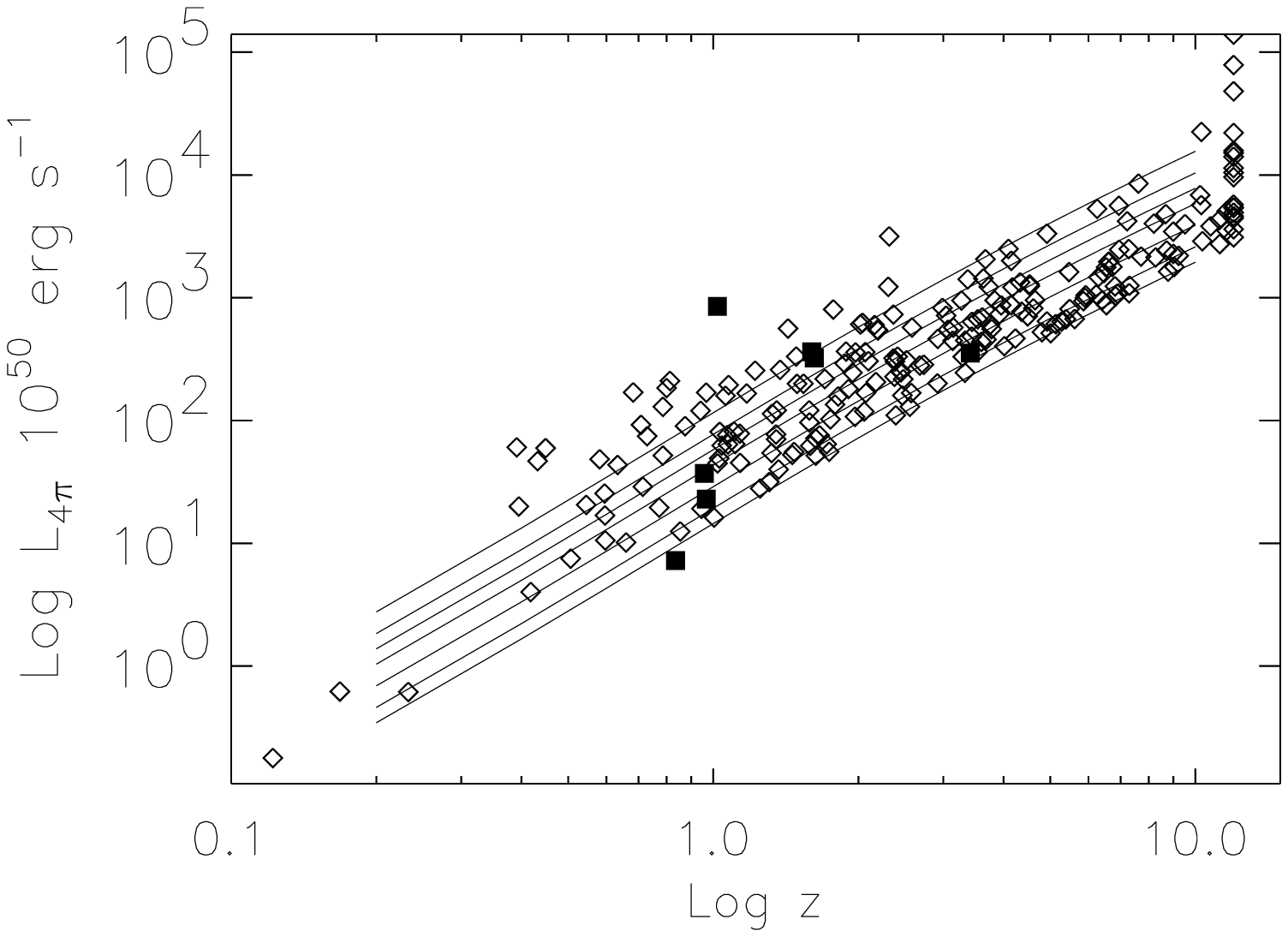]{
Luminosities and redshifts from variability for 220
 long and bright BATSE
GRBs. 
The six solid squares are the GRBs with known redshifts (excluding
GRB980425, see Table 1) that were used to
calibrate the luminosity -- variability relationship.
Bursts with $z > 12$ were placed at $z = 12$.
The solid curves are lines of constant peak flux (photon cm$^{-2}$
s$^{-1}$).
The lower edge of the envelope of events is due to our selection criteria
that $P > 1.5$ photon cm$^{-2}$ s$^{-1}$.
Previous studies only knew the number of GRBs between sets of solid
curves and not their distribution in luminosity and redshift.
The projection of the distribution onto the
luminosity axis would give the GRB luminosity function and the projection
onto the redshift axis would give the progenitor rate if both were 
unaffected by the BATSE threshold. 
\label{lumz}
}

\figcaption[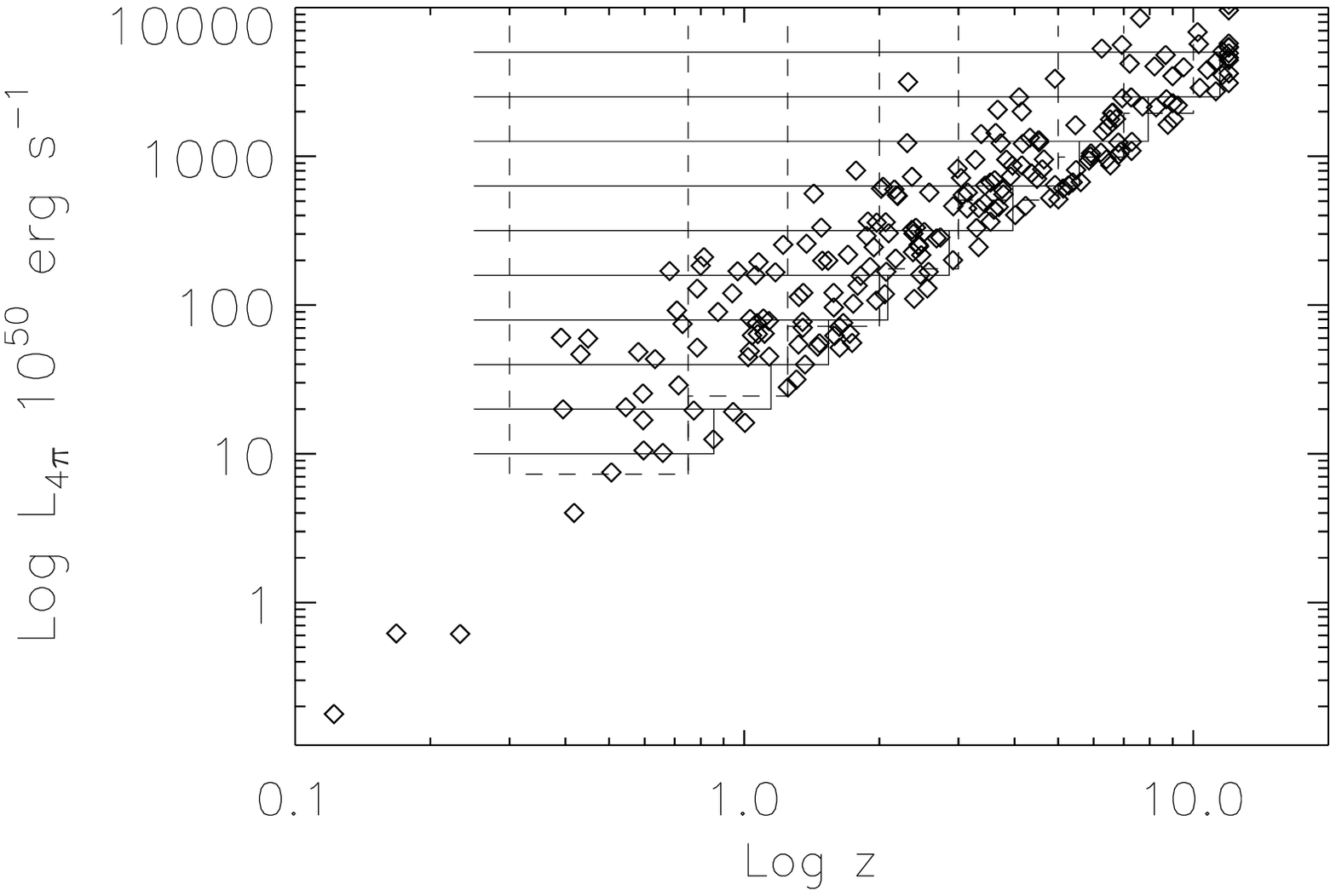]{
Luminosities and redshifts from variability for 220 long and bright BATSE
organized into ranges of redshifts and luminosities. 
The dotted lines
bracket seven regions of the luminosity -- redshift space defined by
ranges of $z$ and where the luminosity function within the range is
unaffected
by the threshold. 
The solid lines
bracket nine regions of the luminosity -- redshift space defined by
ranges of luminosity where the GRB formation rate within the range is
unaffected
by the threshold. 
\label{lumzbound}
}

\figcaption[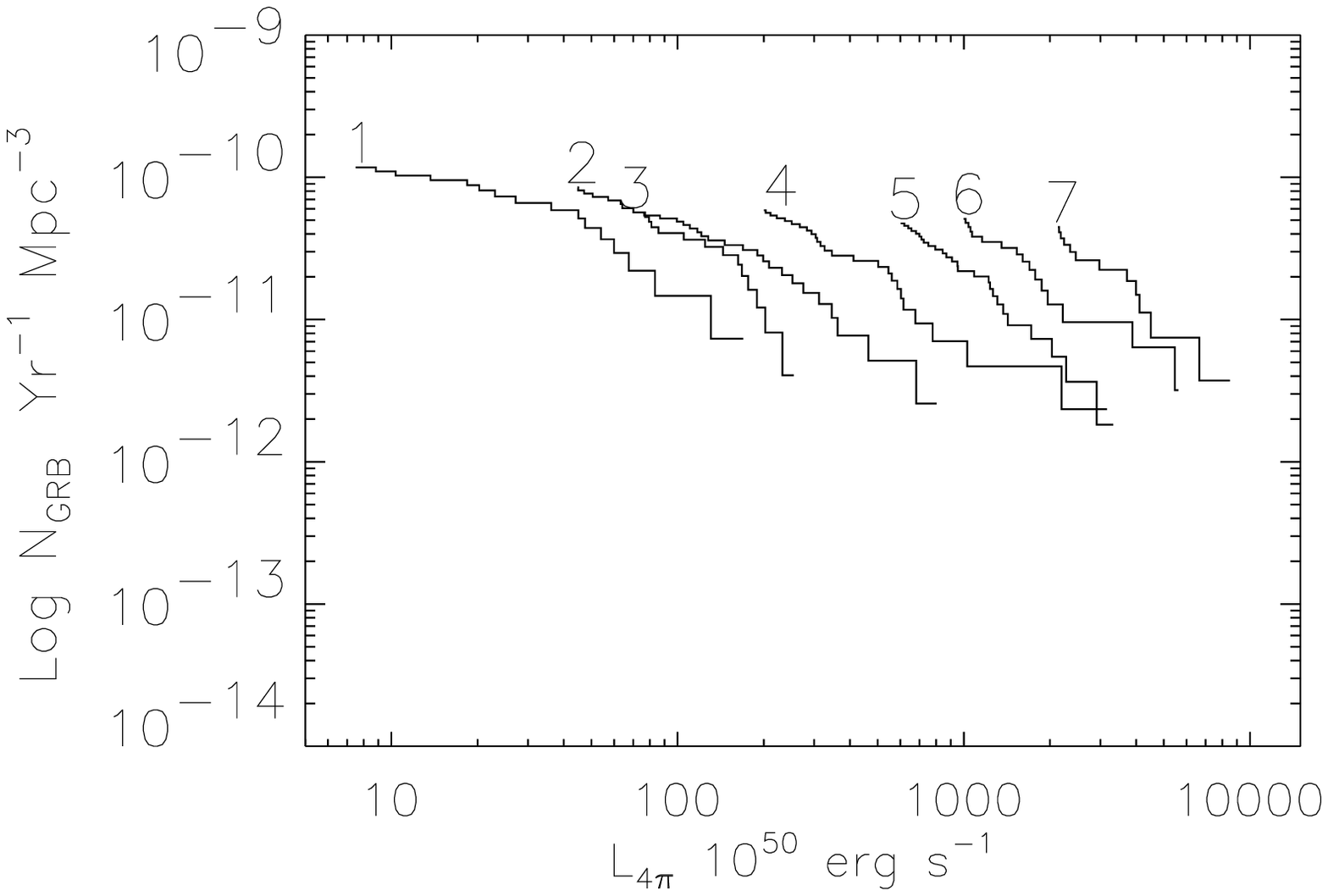]{
The number distribution of GRBs  for
seven ranges of $z$, assuming bursts occur at a constant rate per
time-volume.
In each range, we present the number distribution (\lognl) down to the
luminosity
at which that range is complete.
The seven ranges (labeled 1 to 7) are from Figure
\ref{lumzbound} and are bounded by $z$ = 0.3, 0.75, 1.25,
2.0, 3.0, 5.0, 7.0, and 10.0.
All ranges of $z$ should have the same number distribution, but this is
clearly not the case. We conclude that bursts do not occur at a constant
rate per time-volume.
\label{bfrvolt}
}

\figcaption[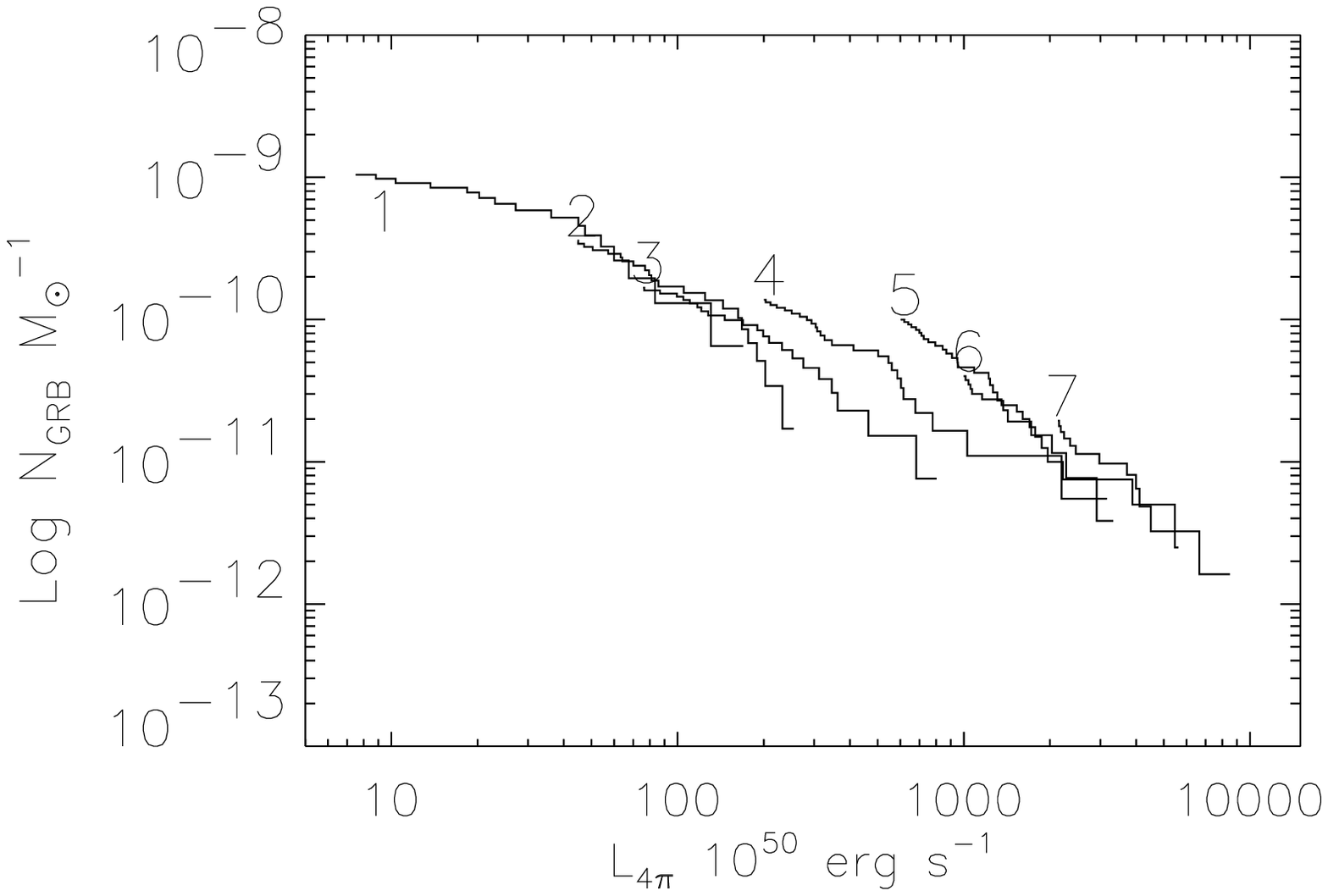]{
The number distribution of GRBs  for
seven ranges of $z$ assuming bursts occur at a rate per
time-volume that is proportional to the SFR of
\cite{steidel}.
In each range, we present the number distribution (\lognl) relative to 
the amount of star formation down to the
luminosity
at which that range is complete.
The $z$ ranges are the same as in Figure \ref{bfrvolt}.
All ranges of $z$ to not fall on a common curve indicating that the
assumption that the burst formation rate follows the star formation rate
is invalid. We conclude there must be relatively more burst formation than star
formation at high $z$.
\label{bfr}
}

\figcaption[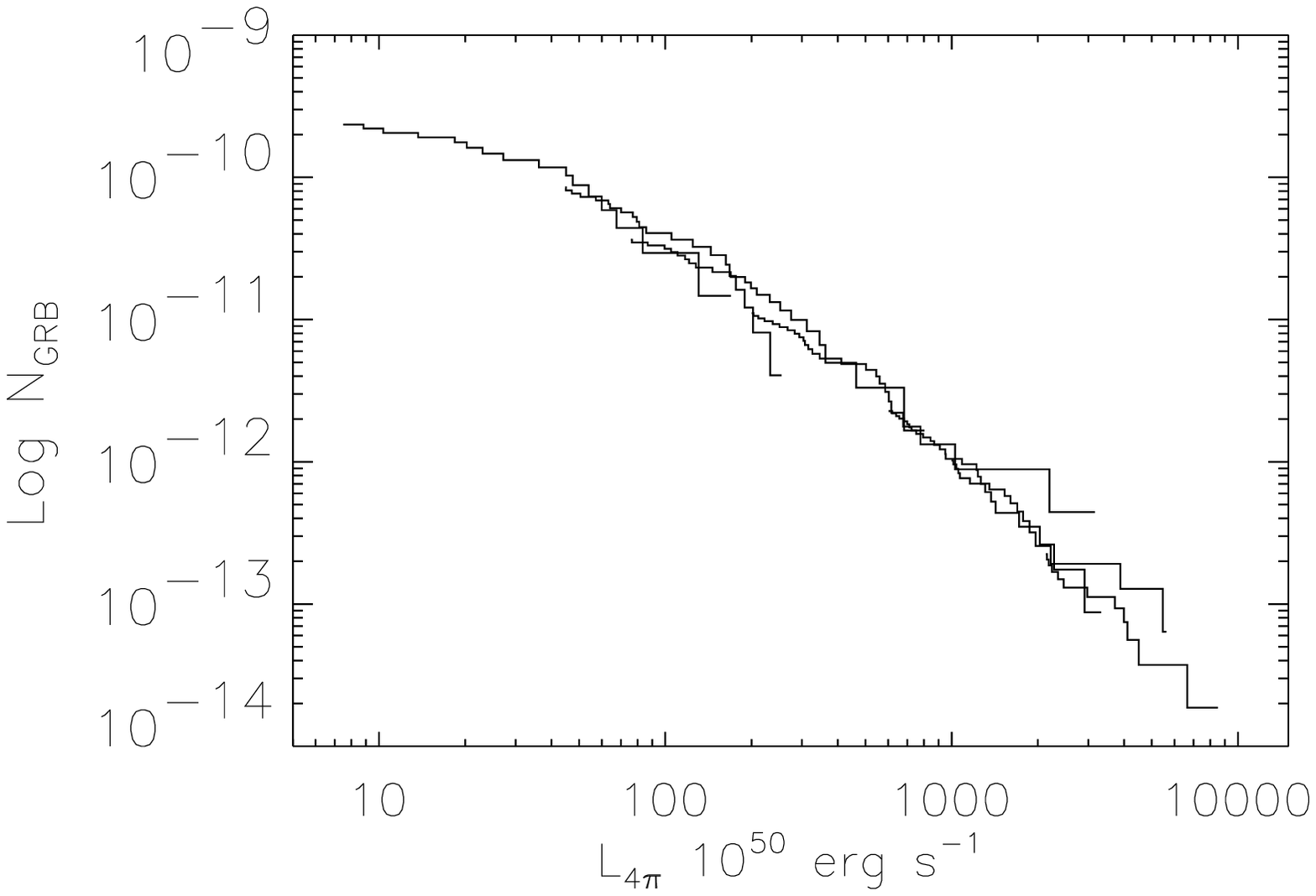]{
The number distribution of GRBs  for
seven ranges of $z$ shifted so that they all fall on a common curve. 
The $z$-ranges are the same as in Figure \ref{bfrvolt}. The amount of
shifting relative to Figure \ref{bfrvolt} gives the relative burst
formation rate.
\label{bfrburstz}
}

\figcaption[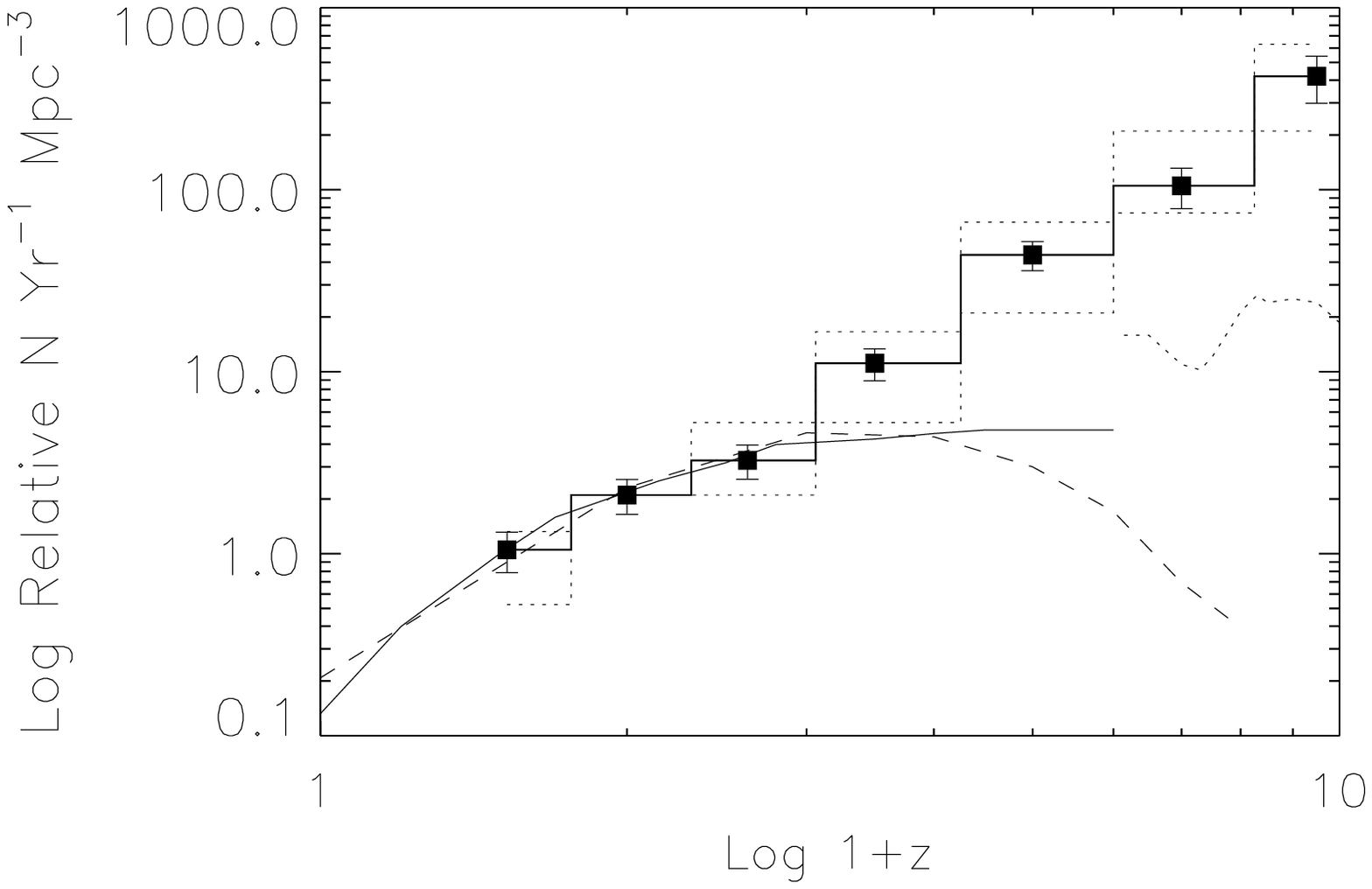]{
The relative GRB formation rate and SFR.
The histogram (with squares) is the GRB formation rate needed to
shift the \lognl\ distributions in 
Figure \ref{bfrvolt} to agree with Figure \ref{bfrburstz}.
The error bars represent the statistical 
uncertainty on each point. This histogram is the burst formation rate determined 
solely from gamma-ray data without assuming a luminosity function or a dependency
on a SFR.
The dotted histograms are based on the two power laws in Figure
\ref{burstzvarfun1} (also see eqs.~[\ref{VTOLUMBOUND},\ref{VTOLUMBOUNDN}])
and they represent the
systematic uncertainty in our burst formation rate.
 The solid
curve is the SFR based on
the observations listed by
\cite{steidel}.
The dotted curve is from the dust-corrected SFR from
observations listed by \cite{blain99}. The SFRs are normalized
to the GRB formation rate at $z$ = 1.
The GRB formation rate scales approximately as $(1+z)^{3.3 \pm 0.3}$.
The  formation rate
for
GRBs continues to rise for $1+z > 2$ whereas the star formation
observations
indicates that the rate levels off or decreases.
The dotted line is a theoretical estimate of the star formation rate at
high z (from \cite{gnedin}).
\label{bfrsfr}
}

\figcaption[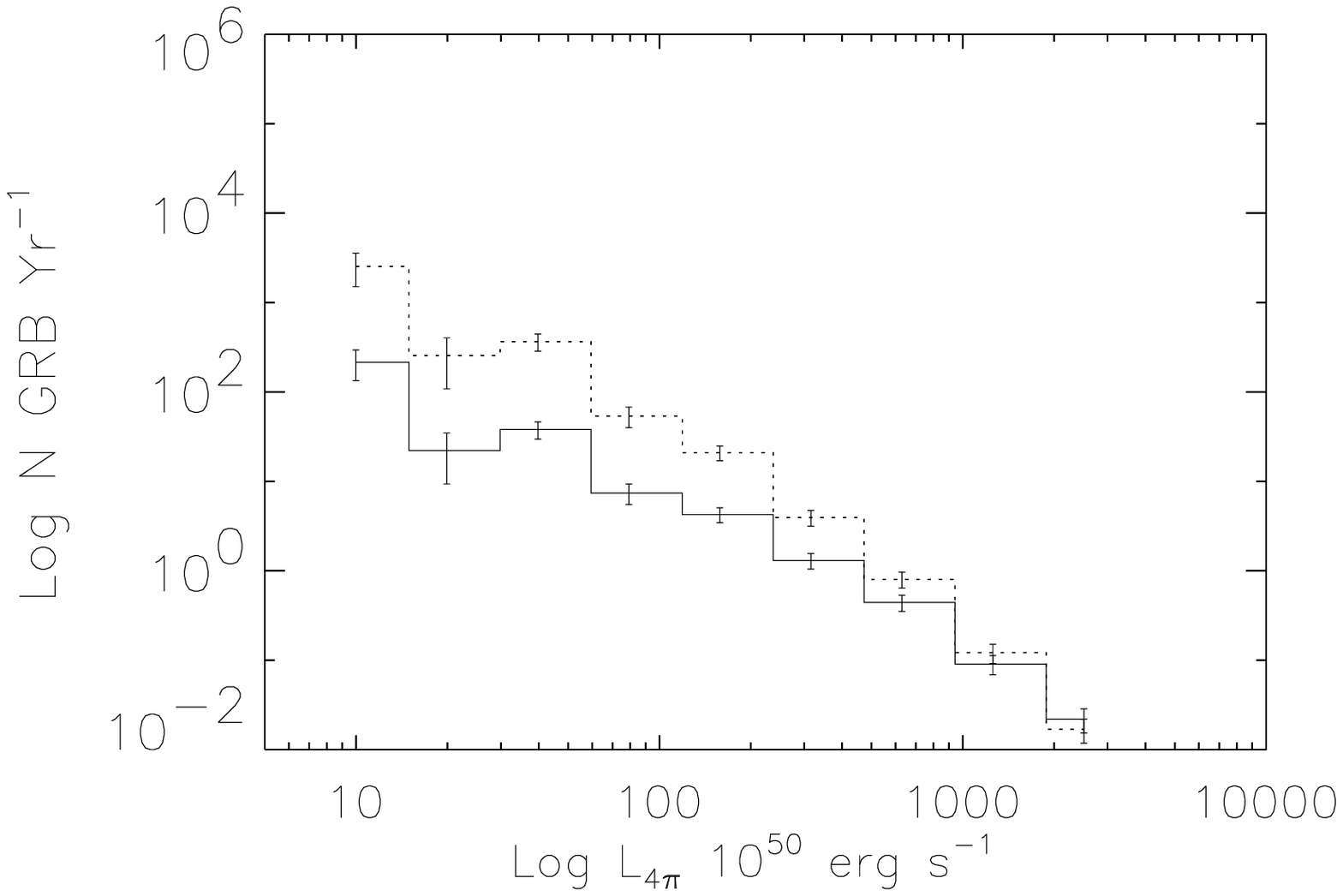]{ 
The GRB luminosity function. The BATSE
threshold prevents us from seeing all GRBs in a particular luminosity
band.  BATSE, at a particular luminosity, is complete to a redshift
$z_c$. We correct the observed number of events from redshifts less than
$z_c$
to the total number that would be seen to $z=\infty$.
The solid line assumes
that the number of GRBs is proportional to the SFR of \cite{steidel}.
The dotted line uses the GRB
formation rate from Figure \ref{bfrsfr} and, thus, is completely derived
from gamma-ray data without any assumptions of star formation rate nor a
form for the GRB luminosity function. 
 The GRB luminosity rolls over at low luminosity, and is a power law
above $4 \times 10^{51}$ erg s$^{-1}$ with
an index of $\sim -1.9$ if based on the SFR and $\sim -2.3$ if based on
the GRB formation rate.
\label{lumfun} }

\figcaption[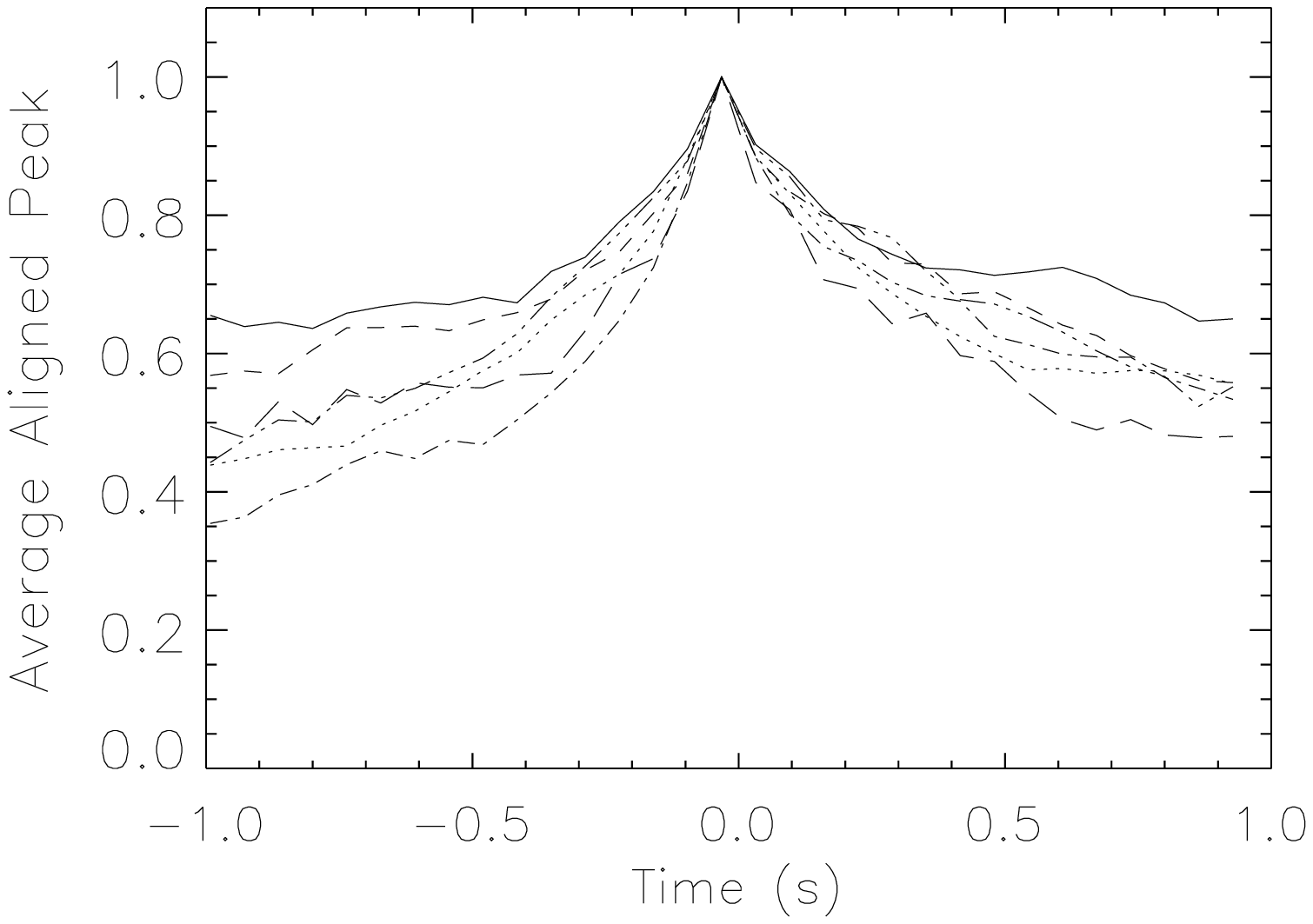]{ 
The aligned peak test for the first 
six ranges of $z$ in Figure \ref{lumzbound}.
The order of the curves (from low to high $z$) is solid,
dotted, dashed, dot-dashed, dot-dot-dashed, and long dashed.
The
largest peak in each 
GRB  within each range of $z$ was aligned and averaged to give the average
peak shape. 
Peaks from lower $z$'s should be wider due to time dilation. 
All ranges give about the
same width, an effect which might be due to the small number of bursts
available.
\label{alignpeakb} }

\figcaption[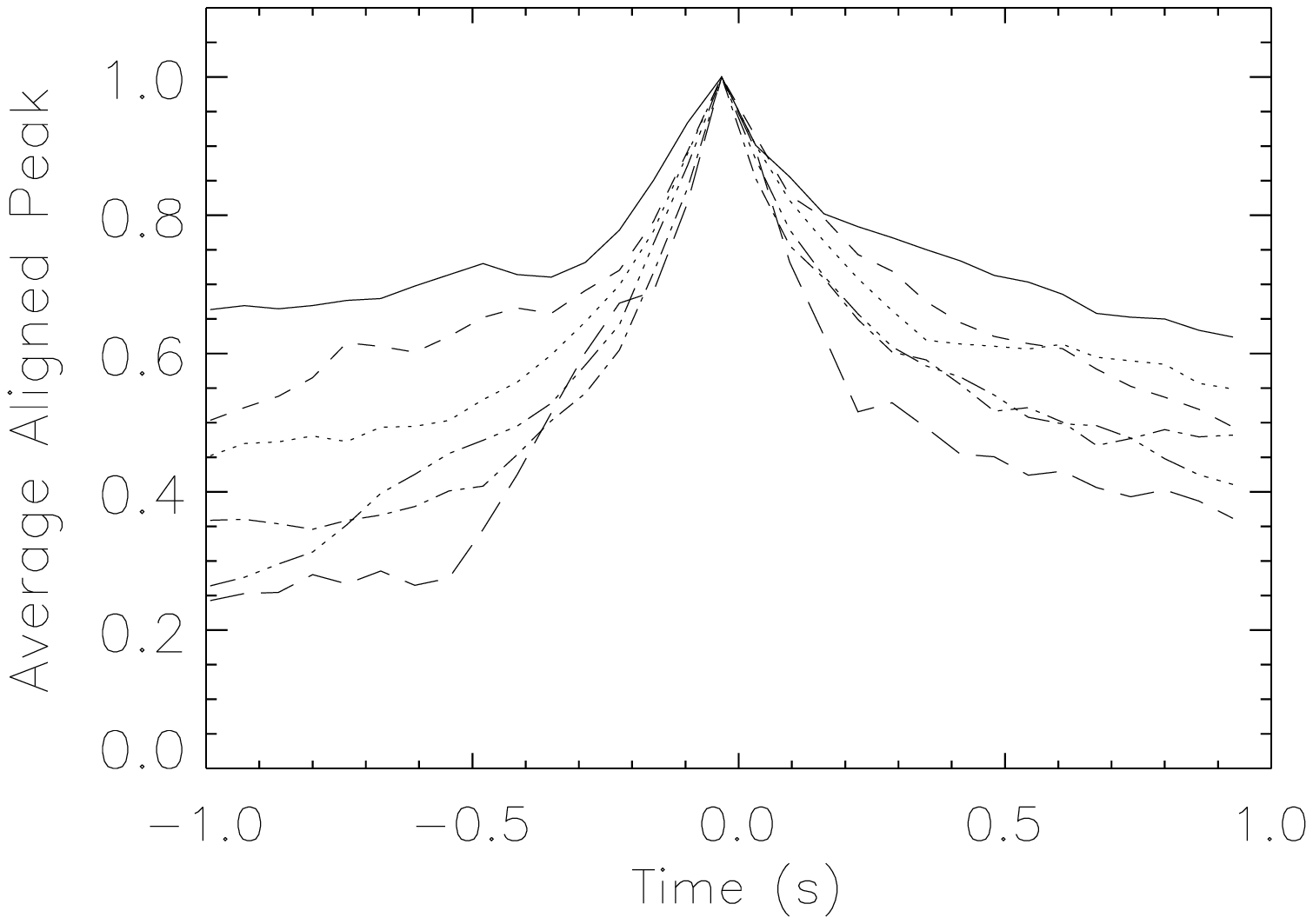]{ 
The aligned peak test for the first six ranges of $z$ in 
Figure \ref{lumzbound}, corrected
for
time
dilation. The
curve definitions are the same as in Figure \ref{alignpeakb}.
The time history of each burst was stretched by the expected time dilation
based on the $z$ estimated from variability (from Table 2). 
Thus, all ranges of $z$ should give
the same peak width. In fact, the dispersion appears to be larger than in
Figure
\ref{alignpeakb}, indicating that this test may not support the idea 
that variability can give
valid redshifts. However, the small number of GRBs and 
limited range of $z$ weakens this
test (see text).
\label{alignpeakz} }



\clearpage
\centerline{Figure \ref{burstzvarfun1}}
\plotone{burstz_varfun1.eps}

\clearpage
\centerline{Figure \ref{burstzvarfun2}}
\plotone{burstz_varfun2.eps}

\clearpage
\centerline{Figure \ref{comparedan}}
\plotone{burstz_compare_dan.eps}

\clearpage
\centerline{Figure \ref{lumz}}
\plotone{burstz_lumz.eps}

\clearpage
\centerline{Figure \ref{lumzbound}}
\plotone{burstz_lumz_bound.eps}

\clearpage
\centerline{Figure \ref{bfrvolt}}
\plotone{burstz_bfr_volt.eps}

\clearpage
\centerline{Figure \ref{bfr}}
\plotone{burstz_bfr.eps}

\clearpage
\centerline{Figure \ref{bfrburstz}}
\plotone{burstz_bfr_burstz.eps}

\clearpage
\centerline{Figure \ref{bfrsfr}}
\plotone{burstz_bfrsfr.eps}

\clearpage
\centerline{Figure \ref{lumfun}}
\plotone{burstz_lumfun.eps}

\clearpage
\centerline{Figure \ref{alignpeakb}}
\plotone{burstz_alignpeak_b.eps}

\clearpage
\centerline{Figure \ref{alignpeakz}}
\plotone{burstz_alignpeak_z.eps}


\clearpage
 
\begin{thebibliography}{}
\newdimen\bibvar \bibvar=-\baselineskip \advance\bibvar-2\parskip


\bibitem[Band et al. 1993]{band93}
Band, D., et al., 1993, ApJ 413, 281.

\bibitem[Blain \& Natarajan (1999)]{blain99n}\vskip\bibvar
\bibitem[Blain \& Natarajan 1999]{blain99}
Blain, A. W., \& Natarajan, P., 1999, submitted to MNRAS,
(astro-ph/9911468)


\bibitem[Bloom et al. 1999]{bloom98}
Bloom, J. S., Kulkarni,  S. R., Harrison, F., Prince, T., Phinney, E. S.
\& Frail, D. A. 1998, ApJ, 506, L105

\bibitem[Bloom, Sigurdsson, \& Pols 1999]{bloom99}
Bloom, J. S., Sigurdsson, S., \& Pols, O. P. 1999, MNRAS, 305, 763

\bibitem[Costa et al.\ 1997]{cos97}
Costa, E., et al., 1997, Nature, 387, 783


\bibitem[Djorgovski 1998]{djorgovski}
Djorgovski, S. G. 1998, ApJ Lett., 508, L17


\bibitem[Fenimore \& Bloom 1995]{fenbloom}
Fenimore, E. E., \& Bloom, J. S. 1995, ApJ, 453, 25

\bibitem[Fenimore \& Ramirez-Ruiz 2000]{fenintext}
Fenimore, E. E., \& Ramirez-Ruiz, E. 2000, submitted to ApJ,
(astro-ph/9909299)


\bibitem[Fenimore et al., 1992]{fenlognp}
Fenimore, E.~E., 1992, Nature, 357, 140


\bibitem[Fenimore et al. (1995)]{autofunn}\vskip\bibvar
\bibitem[Fenimore et al. 1995]{autofun}
Fenimore, E.~E., in't Zand, J.~J., Norris, J.~P., Bonnell, J.~T., \& 
Nemiroff, R.~J., 1995, ApJ 448, L101


\bibitem[Frail et al.\ 1997]{frail97}
Frail, D., Kulkarni, S.~R., Nicastro, L., Feroci, M., \& Taylor, G.~B.,
1997, Nature, 389, 261

\bibitem[Fukugita, Hogan, \& Peebles 1998]{fukugita98}
Fukugita, M., Hogan, C.~J., \& Peebles, P.~J.~E., 1998, ApJ, 508. 518




\bibitem[Galama et al.\ 1999a]{galama980425}
Galama, T.~J., et al., 1999a, Nature, 395, 670

\bibitem[Galama et al.\ 1999b]{galama980425sn} 
Galama, T.~J., et al., 1999b, Astron. Astrophys. Supp. 138, 465

\bibitem[Gehrels et al.\ 2000]{neilswift}
Gehrels, N., et al. 2000,
to appear in the Proc. of the 5th Huntsville
Gamma-ray Burst Symposium.


\bibitem[Gnedin \& Ostriker 1997]{gnedin}
Gnedin, N. Y., \& Ostriker, J. P. 1997, ApJ, 486, 581

\bibitem[Graziani, Lamb, \& Masion 2000]{graziani2000}
Graziani, G., Lamb, D. Q., \& Masion, H., 2000, ApJ, in press
(astro-ph/9810374)


\bibitem[Hogg \& Fruchter 1999]{hogg}
Hogg, D. W., \& Fruchter, A. S., 1999, ApJ 520, 54

\bibitem[Kobayashi, Piran, \& Sari 1997]{ksp97}
Kobayashi, S., Piran, T., \& Sari, R., 1997, ApJ 490, 92

\bibitem[Kommers et al.~(2000)]{kommersn}\vskip\bibvar
\bibitem[Kommers et al.~2000]{kommers}
Kommers, J.~M., et al., 2000, ApJ, 533, 696, astro-ph/9809300

\bibitem[Kulkarni et al.\ 1998]{kulkarni98}
Kulkarni, S., et al. 1998, Nature 393, 35  

\bibitem[Kulkarni et al.\ 1999]{kulkarni0123}
Kulkarni, S., et al. 1999, Nature 398, 389  

\bibitem[Kulkarni et al.\ 2000]{kulkarnihunt}
Kulkarni, S., et al. 2000,
to appear in the Proc. of the 5th Huntsville
Gamma-ray Burst Symposium (astro-ph/0002168)

\bibitem[Lamb 1999]{lambrome}
Lamb, D. Q., 1999, A\&AS, 138 607

\bibitem[Lamb \& Reichart 1999]{lamb99}
Lamb, D. Q., \& Reichart, D. E. 2000, ApJ, in press, (astro-ph/9909002)

\bibitem[Loredo \& Wasserman 1998]{loredo98}
Loredo, T. J., \& Wasserman, I. M. 1998, ApJ, 502, 72

\bibitem[Madau et al. 1996]{madau96}  
Madau, P., et al., 1996, MNRAS, 283, 1388

\bibitem[Metzger et al.\ 1997]{metzger97}
Metzger, M.~R., et al. 1997, Nature, 387, 878

\bibitem[Mitrofanov et al.~1993]{igor93}
Mitrofanov, I. G., et al. 1993, in AIP Conf. Proc. 280, Compton Gamma-Ray
Observatory, ed. M. Friedlander, N. Gehrels, \& D. J. Macomb (New York:
AIP), 761

\bibitem[Norris et al.\ (1994)]{norris94n}\vskip\bibvar
\bibitem[Norris et al.\ 1994]{norris94}
Norris, J.~P., et al., 1994, ApJ 424, 540  
 

 

\bibitem[Norris et al. 1999]{norrishunt}
Norris, J.~P., et al. 1999, Presentation at the 1999
Huntsville GRB conference

\bibitem[Norris et al.\ (2000)]{norris99n}\vskip\bibvar
\bibitem[Norris et al.\ 2000]{norris99}
Norris, J.~P., et al., 2000, ApJ, in press, (astro-ph/9903233)

\bibitem[Odewahn et al.\ 1997]{odewahn97}
Odewahn, S. C., et al. 1997, IAUC 6753

\bibitem[Paciesas et al. 1999]{fourthcat}
Paciesas, et al., 1999, ApJ Supp. 122, 465


\bibitem[Ramirez-Ruiz \& Fenimore (1999)]{enricohuntsn}\vskip\bibvar
\bibitem[Ramirez-Ruiz \& Fenimore 1999]{enricohunts}
Ramirez-Ruiz, E., \& Fenimore, E.~E., 1999, Presentation at the 1999
Huntsville GRB conference

\bibitem[Reichart et al. (2000)]{danvarn}\vskip\bibvar
\bibitem[Reichart et al.~2000]{danvar}
Reichart, D.~E., Lamb, D.~Q., Fenimore, E.~E., Ramirez-Ruiz, E., Cline,
T.~L., \& Hurley, K. 2000, ApJ, submitted, (astro-ph/0004302)

\bibitem[Rowan-Robinson 1999]{rowanrobin}
Rowan-Robinson, M. 1999, Astrophys. \& Space Sci, in press,
(astro-ph/9906308)



\bibitem[Schmidt (1999)]{martinn}\vskip\bibvar
\bibitem[Schmidt 1999]{martin}
Schmidt, M.,  1999, ApJ 523, L117  




\bibitem[Steidel et al.\ 1999]{steidel}
Steidel, C. C., et al. 1999, ApJ, 519, 1


\bibitem[Stern, Poutanen, \& Svennson (1997, 1999)]{sternn}\vskip\bibvar
\bibitem[Stern, Poutanen, \& Svennson 1997]{sterna}
Stern, B., Poutanen, J., \& Svensson, R., 1997, ApJL, 489, L41

\bibitem[Stern, Poutanen, \& Svennson 1999]{sternb}
Stern, B., Poutanen, J., \& Svensson, R., 1999, ApJ, 510, 312


\bibitem[Totani 1997]{totani97}
Totani, T. 1997, ApJ, 486, L71 


\bibitem[Trentham \& Poggiantiin 2000]{trentham00}
Trentham N., \& Poggiantiin, B., in preparation.

\bibitem[van Paradijs et al. 1997]{vanpara97}
van Paradijs, J., et al. 1997, Nature, 386, 686

\bibitem[Vreeswijk et al.\ 1999a]{vreewijka}
Vreeswijk, P., et al.\ 1999a, GCN Cir. 310

\bibitem[Vreeswijk et al.\ 1999b]{vreewijkb}
Vreeswijk, P., et al.\ 1999b, GCN Cir. 496

\bibitem[Wijers et al. 1998]{wijerssfr}
Wijers, R.~A.~M.~J., et al., 1998, MNRAS, 294, L13

\bibitem[Woolsey et al.\ 1999]{woosley980425}
Woolsey, S. E., Eastman, R. G., \& Schmidt, B. P. 1999, ApJ 516, 788


\end{thebibliography}
\end{document}